\documentclass[12pt,aps,prd,onecolumn,superscriptaddress,nofootinbib]{revtex4-2}

\usepackage{amsmath,amssymb,amsfonts,mathrsfs,bbm,latexsym}
\usepackage{graphicx}
\usepackage[english]{babel}
\usepackage{multirow}
\usepackage{url}
\usepackage{slashed}

\usepackage{cancel}

\usepackage{orcidlink}

\usepackage[normalem]{ulem}

\newcommand{\be}{\begin{equation}}
\newcommand{\ee}{\end{equation}}
\newcommand{\ba}{\begin{array}}
\newcommand{\ea}{\end{array}}
\newcommand{\bea}{\begin{eqnarray}}
\newcommand{\eea}{\end{eqnarray}}

\begin{document}

\title{\small Freeze-in Production of Non-Abelian Millicharged Vector Dark Matter}

\author{Van Que Tran \orcidlink{0000-0003-4643-4050}}
\email{vqtran@phys.ncts.ntu.edu.tw} 
\affiliation{Physics Division, National Center for Theoretical Sciences, National Taiwan University, Taipei 106319, Taiwan}
\affiliation{Phenikaa Institute for Advanced Study, Phenikaa University, Nguyen Trac, Duong Noi, Hanoi 100000, Vietnam}

\author{Tzu-Chiang Yuan \orcidlink{0000-0001-8546-5031}}
\email{tcyuan@phys.sinica.edu.tw}
\affiliation{Institute of Physics, Academia Sinica, Nangang, Taipei 11529, Taiwan}

\begin{abstract}

We present the first predictive realization, to our knowledge, of vector freeze-in dark matter from a hidden non-Abelian \(SU(2)\) gauge sector spontaneously broken by a Higgs triplet to a residual $U(1)$ symmetry with a massless dark photon mediator. A massive dark vector particle-antiparticle pair acquires small millicharges through a dimension-4 kinetic-mixing term with an induced coefficient $\epsilon$, generated by an effective dimension-5 operator involved the Higgs triplet.  The dark sector interactions are governed by the hidden gauge coupling $g_D$, providing a weak connection to the Standard Model to realize the freeze-in dark matter production mechanism. Solving the two-temperature Boltzmann evolution including plasmon decay, we find a wide region of parameter space consistent with the observed relic abundance while satisfying astrophysical and cosmological constraints. This minimal framework connects non-Abelian vector dynamics with long-range dark forces and may be probed by upcoming sub-GeV dark matter direct-detection experiments.

\end{abstract}

\maketitle

\newpage


\section{Introduction}
\label{sec:intro}

The absence of dark matter (DM) candidates within the Standard Model (SM) strongly motivates the exploration of new particles and interactions beyond the SM. Among the possibilities, dark sectors equipped with their own gauge symmetries--potentially weakly coupled to the visible sector--offer a natural framework for dark matter and long-range interactions. In particular, hidden vector bosons charged under non-Abelian gauge groups are well motivated in ultraviolet-complete theories and may provide a rich phenomenology.

One intriguing possibility is that hidden-sector particles carry a small electric millicharge, allowing them to interact feebly with SM photons. While all observed SM particles carry charges quantized in units of \(e/3\), millicharged particles (MCPs) naturally arise when a hidden \(U(1)_D\) mixes kinetically with the SM hypercharge \(U(1)_Y\). This idea was first proposed by Holdom~\cite{Holdom:1985ag}, who showed that gauge-invariant kinetic-mixing between two \(U(1)\) fields is allowed at the renormalizable level and generically induces small effective electric charges for otherwise neutral hidden-sector matter.

Since then, millicharged particles have been studied extensively, both as dark matter candidates and as probes of hidden sectors. MCPs are often modeled as fermions with fractional electric charge \(\epsilon e\), where \(\epsilon\) denotes the charge fraction. Such a small charge may arise from kinetic mixing between \(U(1)_Y\) and a hidden \(U(1)_D\), resulting either in a massless dark photon~\cite{Holdom:1985ag} or a massive one via the Stueckelberg mechanism~\cite{Feldman:2007wj,Cheung:2007ut}. Numerous experimental efforts have searched for MCPs through collider and beam-dump experiments, including LEP~\cite{OPAL:1995uwx}, SLAC~\cite{Prinz:1998ua}, ArgoNeuT~\cite{ArgoNeuT:2019ckq}, the MilliQan pathfinder at the LHC~\cite{Ball:2020dnx}, and various neutrino experiments~\cite{Magill:2018tbb,Marocco:2020dqu}. These searches constrain \(\epsilon\) to lie between \( \mathcal{O}(10^{-4}) \) and \( \mathcal{O}(10^{-1}) \) for MCP masses in the MeV–TeV range (see~\cite{Fabbrichesi:2020wbt} for a review).

In the low-mass regime (\( m_{\rm MCP} \lesssim \text{keV--GeV} \)), MCPs are subject to powerful constraints from astrophysical and cosmological observations. Plasmon decay in red giants, white dwarfs, and horizontal branch stars places tight limits on MCPs with \(\epsilon \lesssim 10^{-14}\)~\cite{Davidson:2000hf,Vogel:2013raa,Raffelt:1996wa}. Additional bounds arise from supernova cooling, cosmic microwave background (CMB) anisotropies, and baryon-coupled drag effects that modify the 21-cm signal~\cite{Dubovsky:2003yn,McDermott:2010pa,Kovetz:2018zes}. Cosmic-ray production also offers complementary constraints~\cite{Plestid:2020kdm}.

Most studies to date have focused on Abelian hidden sectors, where the MCP originates from kinetic mixing between \(U(1)_Y\) and a hidden \(U(1)_D\). However, non-Abelian dark sectors can give rise to qualitatively new effects, particularly when spontaneously broken to an unbroken Abelian subgroup. In such models, effective millicharges can emerge via higher-dimensional operators, especially in the presence of scalar fields charged under the hidden symmetry. In addition, non-Abelian vector dark matter can be naturally implemented as self-interacting dark matter due to its non-Abelian nature~\cite{Tran:2023lzv,Ko:2020qlt}.

In this work, we study a scenario in which dark matter arises from a hidden \(SU(2)_D\) gauge theory, spontaneously broken to a residual \(U(1)_D\) by a real Higgs triplet. The symmetry breaking gives rise to a massless dark gauge boson, referred to as the dark \(Z\)~\footnote{Owing to its masslessness, it may also be referred to as the dark photon. Both terms, dark photon and dark \( Z \), are used interchangeably throughout this article.}. The extended gauge symmetry also enables a dimension-5 effective operator involving the hidden Higgs triplet and the field strengths of \(SU(2)_D\) and SM hypercharge \(U(1)_Y\). This dimension-5 operator further induces a dimension-4 kinetic-mixing operator between \(U(1)_Y\) and the neutral component of the broken \(SU(2)_D\), thereby generating a small electric millicharge for the massive dark gauge bosons, which serves as a stable vector dark matter candidate.

Millicharged particles can contribute to the dark matter relic density if they are stable. However, strong constraints from CMB observations either require MCPs to decouple from the SM plasma before recombination or limit them to comprise only a subdominant component of the dark matter~\cite{McDermott:2010pa,Dolgov:2013una}. In the former case, a freeze-in production mechanism is necessary to generate the correct relic density without violating cosmological bounds~\cite{Dvorkin:2019zdi, Jana:2024iig}.

We analyze in detail the cosmological and astrophysical implications of this framework. The relic abundance of the vector dark matter is computed via the freeze-in mechanism, incorporating a two-temperature Boltzmann evolution~\cite{Foot:2014uba,Foot:2016wvj,Hambye:2019dwd} for the visible and hidden sectors. We include in-medium effects such as plasmon decay~\cite{Braaten:1993jw,Dvorkin:2019zdi}.

The remainder of this manuscript is structured as follows. 
In Sec.~\ref{sec:ModelSetUp}, we present the model under study. We develop a simple extension of the SM by introducing a hidden \(SU(2)_D\) gauge group, which is spontaneously broken to a massless \(U(1)_D\) via a real Higgs triplet. 
We explain how this setup naturally gives rise to millicharged vector dark matter through Abelian--non-Abelian kinetic mixing, previously discussed in the context of Standard Model effective field theory (SMEFT)~\footnote{For a recent review on SMEFT, see, for example,~\cite{Aebischer:2025qhh}.} and its SMEFT+triplet extension~\cite{Tran:2024srm}. 
It is then followed by 
a detailed discussion of the rotation angle independence and ``millicharged basis'' in Sec.~\ref{sec:milichargedbasis}. 
We discuss the freeze-in dark matter production in the model in 
Sec.~\ref{sec:FreezeIn}. 
Section~\ref{sec:Constraints} is devoted to various constraints from astrophysics, cosmology, and direct detection experiments. 
We examine stringent constraints on a massless dark photon from the ellipticity profile of galaxy NGC720, as well as from the effective number of relativistic species inferred from the CMB and big bang nucleosynthesis (BBN).
Our numerical results are presented in Sec.~\ref{sec:NumericalResults}, and we conclude in Sec.~\ref{sec:Conclusions}.

Some mathematical details are deferred to the Appendices. Appendix~\ref{app:UVcompletion} presents a simple UV completion for the dimension-5 effective operator involving the hidden gauge group $SU(2)_D$, the hidden triplet Higgs, and the SM hypercharge group $U(1)_Y$. Appendix~\ref{app:dxidT} derives the evolution equation for the temperature ratio $\xi = T_h/T$ between the hidden and visible sectors. Appendix~\ref{app:rho_p_s} collects the expressions for the pressure, energy density, and entropy density in both sectors. Appendix~\ref{app:Jh} lists the relevant source terms entering the coupled Boltzmann equations. Appendix~\ref{app:CoupledBoltzmann} presents the coupled Boltzmann equations for the visible and hidden sectors. Appendix~\ref{app:xsec_decay} compiles the cross sections and decay rates relevant for freeze-in production, while plasmon decay into millicharged vector dark matter is discussed separately in Appendix~\ref{app:plasmondecay}.


\section{Model Setup}
\label{sec:ModelSetUp}

\subsection{Lagrangian and Abelian--non-Abelian kinetic mixing} 
We consider a dark sector with an $SU(2)_D$ gauge symmetry and a real scalar triplet $\Sigma = \Sigma^a T_D^a$, where $T_D^a = \sigma^a / 2 \, (a=1,2,3)$ are the generators in the fundamental representation.
The Lagrangian is
\begin{equation}
\label{eq:lag}
\mathcal{L}=\mathcal{L}_{\rm SM}+\mathcal{L}_{\rm DS}+\mathcal{L}_{\rm mix}- \Delta V(\Phi,\Sigma) \; ,
\end{equation}
where $\mathcal{L}_{\rm SM}$ is the SM part, 
\begin{align}
\mathcal{L}_{\rm DS}&=-\tfrac{1}{4}W^a_{D\mu\nu}W^{a\mu\nu}_D -\tfrac{1}{2}\,\mathrm{Tr}\!\left[(D_\mu\Sigma)(D^\mu\Sigma)\right] \; , \\
D_\mu\Sigma&=\partial_\mu\Sigma - i g_D [W^a_{D\mu}T^a_D,\Sigma] \; ,
\end{align}
and $\Delta V(\Phi,\Sigma)$ (to be defined shortly) is  the scalar potential that includes the real triplet and a portal term involving the SM Higgs $\Phi$ and $\Sigma$.
Here $W^a_{D\mu} \, (W^a_{D\mu\nu})$ is the 
gauge field (field strength), $g_D$ its gauge coupling, and $\mathcal{L}_{\rm mix}$ is the kinetic-mixing term.

An Abelian–non-Abelian kinetic-mixing between the hidden $SU(2)_D$ and SM $U(1)_Y$ arises from the dimension-5 operator~\footnote{A similar operator linking the SM electroweak gauge group  $SU(2)_L \times U(1)_Y$ by a weak triplet Higgs of zero hypercharge was analyzed in~\cite{Tran:2024srm}.}
\begin{equation}
\label{eq:O5}
{\cal O}_5 = \frac{c_5 g_D g'}{\Lambda} \, \mathrm{Tr}\!\left[ \Sigma W_D^{\mu\nu} \right] B_{\mu\nu} \;,
\end{equation}
with $c_5$ dimensionless and $g'$ the hypercharge coupling. 
The dimension-5 operator in~(\ref{eq:O5}) can be induced, for example, by integrating out a heavy Dirac fermion \( \Psi(\mathbf{1}, \mathbf{2}, y) \), where the numbers in parentheses denote its representations under \( SU(2)_L \times SU(2)_D \times U(1)_Y \), running in a triangle loop diagram. The explicit one-loop matching and a brief discussion of phenomenological constraints on the heavy fermion are provided in Appendix~\ref{app:UVcompletion}.

When $\Sigma$ acquires a VEV, $\langle \Sigma \rangle = \tfrac{1}{2}\,\mathrm{Diag}(v_\Sigma,-v_\Sigma)$, $SU(2)_D$ breaks to $U(1)_D$ and~(\ref{eq:O5}) induces
\begin{equation}
\mathcal{L}_{\rm mix} \supset -\tfrac{\epsilon}{2}\,W^{3\mu\nu}_{D}B_{\mu\nu} \; , \qquad 
\epsilon = -\frac{c_5 g_D g' v_\Sigma}{\Lambda} \; .
\end{equation}
We note that the kinetic-mixing parameter $\epsilon$ arises naturally from the dimension-5 operator generated by heavy fermions, with loop suppression and heavy mass scaling ensuring its smallness.

\subsection{Scalar potential and mixing}
\label{subsec:ScalarPotential}
The triplet scalar field can be conveniently expressed as a matrix
 \be
 \Sigma =  \Sigma^a \frac{\sigma^a }{2} = \frac{1}{2}
 \begin{pmatrix}
  v_\Sigma + \Sigma^0  & \sqrt{2} \Sigma^p \\  
  \sqrt{2} \Sigma^m & - v_\Sigma - \Sigma^0\\  
  \end{pmatrix},
  \label{eq:tripletSigma}
  \ee
where $\sigma^a (a=1,2,3)$ are the Pauli's matrices, $\Sigma^0 = \Sigma^3$ and $\Sigma^{p,m} = \frac{1}{\sqrt 2} ( \Sigma^1 \mp i\Sigma^2)$ are the neutral and dark charged triplet scalar bosons respectively.

In standard notation, the SM doublet \( \Phi \) is parametrized as
\be
\Phi = \begin{pmatrix}
    G^+ \\
    \frac{1}{\sqrt{2}}( v + \phi + i G^0) 
\end{pmatrix} \; ,
\ee
with \( v \) the SM Higgs VEV. 

The scalar potential in the model is given by
\bea
\label{eq:scalarpot}
V(\Phi, \Sigma) & = & - \mu^2_\Phi (\Phi^\dagger\Phi) + \lambda_\Phi (\Phi^\dagger\Phi)^2  + \Delta V(\Phi, \Sigma) \; , \\
\Delta V(\Phi, \Sigma) & = &
- \frac{1}{2} \mu_\Sigma^2  (\Sigma^a \Sigma^a) + \frac{1}{4}\lambda_\Sigma (\Sigma^a \Sigma^a)^2 + \frac{1}{2}\lambda_{\Phi\Sigma} (\Phi^\dagger\Phi) (\Sigma^a \Sigma^a) \; . 
\eea
Combining the requirements of validity of perturbative calculations, tree-level perturbative unitarity in the scalar sector, and bounded from below for the scalar potential leads us to the following theoretical constraints
\bea
0  \; < & \lambda_{\Phi} , \lambda_\Sigma & \leq \frac{4 \pi}{3} \; , \\
-2 \sqrt{\lambda_\Phi \lambda_\Sigma} \;  < & \lambda_{\Phi \Sigma} & \leq 8 \pi \; .
\label{eq:PotentialConstraints}
\eea

The minimization conditions, 
$\partial V(\Phi,\Sigma) /\partial v = 0$ and
$\partial V(\Phi,\Sigma) /\partial v_\Sigma = 0$,
lead to
\begin{align}
\mu_\Phi^2 &=
\lambda_\Phi v^2
+\frac12\lambda_{\Phi\Sigma} v_\Sigma^2 \; ,
\label{eq:min_cond_phi}\\
\mu_\Sigma^2 &=
\lambda_\Sigma v_\Sigma^2
+\frac12\lambda_{\Phi\Sigma} v^2 \; .
\label{eq:min_cond_sigma}
\end{align}
The $CP$-even neutral scalar fields $(\phi,\Sigma^0)$ mix after symmetry breaking.
Expanding the scalar potential to quadratic order and applying the minimization
conditions~\eqref{eq:min_cond_phi}--\eqref{eq:min_cond_sigma},
the mass-squared matrix in the basis $(\phi,\Sigma^0)$ is
\begin{equation}
\mathcal{M}^2_{\text{scalar}} =
\begin{pmatrix}
2\lambda_\Phi v^2 & \lambda_{\Phi\Sigma} v v_\Sigma \\
\lambda_{\Phi\Sigma} v v_\Sigma & 2\lambda_\Sigma v_\Sigma^2
\end{pmatrix}.
\label{eq:mass_matrix_even}
\end{equation}
The matrix can be diagonalized by an orthogonal transformation,
\begin{align} 
\begin{pmatrix} 
h \\ h_D
\end{pmatrix}
= 
\begin{pmatrix} 
\cos\beta & -\sin\beta \\
\sin\beta & \cos\beta
\end{pmatrix}
\begin{pmatrix} 
\phi \\ \Sigma^0
\end{pmatrix}\; ,
\end{align}
where the mixing angle satisfies
\begin{equation}
\tan 2\beta =
\frac{\lambda_{\Phi\Sigma} v v_\Sigma}
{\lambda_\Phi v^2 - \lambda_\Sigma v_\Sigma^2} \; .
\label{eq:mixing_angle_beta}
\end{equation}
The physical $CP$-even scalar masses are
\begin{align}
m_{h,h_D}^2
&=
\lambda_\Phi v^2 + \lambda_\Sigma v_\Sigma^2
\mp
\sqrt{
(\lambda_\Phi v^2 - \lambda_\Sigma v_\Sigma^2)^2
+ (\lambda_{\Phi\Sigma} v v_\Sigma)^2
} \; .
\label{eq:scalar_masses}
\end{align}
We identify $h$ as the Higgs boson observed at the LHC with a mass of $125$ GeV, while $h_D$ is an extra scalar. 
The Higgs data measurements at the LHC constrain the mixing angle $|\sin\beta| \lesssim 0.2$~\cite{ATLAS:2022vkf}. 
On the other hand, the charged components $\Sigma^{p,m}$ acquire zero masses and serve as the unphysical Goldstone bosons that are absorbed by the longitudinal components $W_L^{p,m}$ of the hidden gauge bosons, which constitute the vector dark matter candidates in this model.

\subsection{Gauge Bosons}
\label{subsec:GBs}

The kinetic energy Lagrangian for the gauge sector in the model can be expressed as
 \be
 \label{eq:kin}
 {\cal L}_{\rm kin} =  -\frac{1}{4} {B}_{\mu\nu} {B}^{\mu\nu} - \frac{1}{4} {W}^a_{\mu\nu} {W}^{a\mu\nu} - \frac{1}{4} {W}^{a}_{D\mu\nu} {W}_D^{a \mu\nu} 
 - \frac{1}{2} \epsilon {W}_D^{3 \mu\nu} {B}_{\mu\nu} \; .
 \ee
 To eliminate the kinetic-mixing term in Eq.~(\ref{eq:kin}), the following field redefinition can be applied:
\be
\label{eq:fieldredefinition2}
\begin{pmatrix}
    {W}^3_{D \mu} \\
    {B}_\mu 
\end{pmatrix}
    \to  
\begin{pmatrix}
\frac{1}{\sqrt{1-\epsilon^2}} & 0\\
-\frac{\epsilon}{\sqrt{1-\epsilon^2}} & 1   
\end{pmatrix}
\cdot
\begin{pmatrix}
c_\alpha & -s_\alpha\\
s_\alpha & c_\alpha   
\end{pmatrix}
\cdot
\begin{pmatrix}
    {W}^3_{D \mu} \\
    {B}_\mu 
\end{pmatrix},
\ee
where $c_\alpha \equiv \cos \alpha$, $s_\alpha \equiv  \sin \alpha$ and the angle $\alpha$ is arbitrary. 
We emphasize that, since both the photon and dark photon are massless as we will demonstrate below, the angle $\alpha$ reflects a redundancy in the choice of basis in the unbroken Abelian subgroup $U(1)_Y \times U(1)_D$. Physical observables must be independent of $\alpha$, although intermediate quantities such as gauge interaction vertices and therefore individual amplitude may exhibit explicit $\alpha$-dependence. We will demonstrate this explicitly in Sec.~\ref{sec:milichargedbasis}.

The redefinition in Eq.~(\ref{eq:fieldredefinition2}) affects the covariant derivative, which now takes the form
\bea
\begin{aligned}
\label{eq:covariantD}
D_{\mu} & =  \partial_\mu \cdot \mathbbm{1} - i g W^{a}_\mu   T^a 
- i B_\mu \left( \tilde{g}' \, Y - 
   \frac{s_\alpha g_D}{\sqrt{1-\epsilon^2}}   T_D^3 \right)
\\ 
& - i g_D \left(W^1_{D\mu}  T_D^1+  W^2_{D\mu}  T_D^2\right)  
-i W^3_{D\mu} \left(  \tilde{g}{''}  Y + 
   \frac{c_\alpha g_D}{\sqrt{1-\epsilon^2}}  T_D^3  \right) \; ,  
\end{aligned}
\eea
where 
\bea
\label{eq:gtp}
\tilde{g}' &=& g' \left(c_\alpha + \frac{\epsilon}{\sqrt{1-\epsilon^2}} s_\alpha \right),
\\
\label{eq:gtpp}
\tilde{g}{''} &=& g' \left(s_\alpha - \frac{\epsilon}{\sqrt{1-\epsilon^2}} c_\alpha \right).
\eea
Here
$T^a$ ($T^a_D$) with $a=1,2,3$ are the $SU(2)_L$ ($SU(2)_D$) generators, $Y$ is the hypercharge and $g'$, $g$ and $g_D$ are gauge couplings of the $U(1)_Y$, $SU(2)_L$ and $SU(2)_D$, respectively. For doublet, $(T^a_{(D)})_{ij}= (\tau^a)_{ij}/2$ with $i,j=1,2$; while for triplet $(T^a_{(D)})_{bc} = - i \epsilon_{abc}$ with $a,b,c=1,2,3$.

The covariant derivatives of the $SU(2)_L$ doublet scalar $\Phi$ and the $SU(2)_D$ triplet $\Sigma$ can then be written down as 
\bea
\label{eq:covPhi}
D_{\mu} \Phi &=& \left( \partial_\mu \cdot \mathbbm{1} - i g W_\mu^a \frac{\tau^a}{2} - i \frac{1}{2}\tilde{g}' B_\mu - i \frac{1}{2} \tilde{g}''  W^3_{D\mu} \right)\Phi \;, \\
\label{eq:covSigma}
D_{\mu} \Sigma^a&=& \biggl[ \delta^{ab} \partial_\mu  - g_D \left(W^1_{D\mu}  \epsilon_{1ab}+  W^2_{D\mu}  \epsilon_{2ab} \right) 
+ \frac{ g_D}{\sqrt{1-\epsilon^2}} \left( s_\alpha B_\mu - c_\alpha W^3_{D\mu} \right) \epsilon_{3ab}
\biggr] \Sigma^b \; .
\eea

The gauge boson mass terms come from the squares of Eqs.~(\ref{eq:covPhi}) and~(\ref{eq:covSigma}), evaluated at the scalar fields' VEVs.
The mass terms are then given as
\be
\label{eq:gaugemass}
{\cal L}^{\rm gauge}_{\rm mass} =  \frac{1}{4} g^2 v^2 W_\mu^+ W_\mu^- + \frac{1}{8} v^2 \left(- g W^3_\mu + \tilde{g}' B_\mu + \tilde{g}''  W^3_{D\mu} \right)^2  +  g_D^2 v_\Sigma^2 W_{\mu}^p W_{\mu}^m \; ,
\ee
where we define $W_{\mu}^\pm = \frac{1}{\sqrt{2}} (W_{\mu}^1 \mp i W_{\mu}^2)$, $W_{\mu}^{(p,m)} = \frac{1}{\sqrt{2}} (W_{D\mu}^1 \mp i W_{D\mu}^2)$ and $v = 246$ GeV. 
The mass of the SM gauge boson $W^\pm$ remains the same as $m_{W^\pm} = g v / 2$, while the dark charged vector boson $W^{p,m}$ has a mass $m_{W^{p,m}} \equiv m_{W'} = g_D v_\Sigma$. 

In the basis of $V = (B, W^3, W^3_D)$, the second term in Eq.~(\ref{eq:gaugemass}) yields the following symmetric mass matrix for the neutral gauge bosons:
\be
M_{1}^2 = \frac{1}{4} v^2
\begin{pmatrix} 
\tilde{g}'^{2} & - \tilde{g}' g &\tilde{g}' \tilde{g}'' \\
  - \tilde{g}' g &  g^2 &  -\tilde{g}'' g \\
 \tilde{g}' \tilde{g}'' & -\tilde{g}'' g &\tilde{g}''^2
\end{pmatrix}. 
\label{eq:massmatrix_massless}
\ee
It is straightforward to verify that the mass matrix in Eq.~(\ref{eq:massmatrix_massless}) has two zero eigenvalues, corresponding to the massless photon and the massless dark photon (or dark \( Z \)), and one nonzero eigenvalue
\be
\label{eq:Zbosonmass}
\begin{aligned}
m_{Z}^2 & = \frac{1}{4} v^2 \left(g^2 + \tilde{g}'^{2} + \tilde{g}''^2 \right) =  \frac{1}{4} v^2 \left( g^2 +  \bar{g}'^2 \right), \\
\mathrm{ with } \qquad  \bar{g}'^2 & =  \tilde{g}'^{2} + \tilde{g}''^2 = \frac{g'^2}{1-\epsilon^2} \; ,
\end{aligned}
\ee
which corresponds to the mass of the SM $Z$ boson.  Since $\bar{g}'$ does not depend on the rotation angle $\alpha$, the $Z$ boson mass is also independent of $\alpha$.

An orthogonal transformation $\mathcal{O}$ can be used to rotate the gauge basis $V_i$ into the mass basis $E_i = (A, Z, Z')$, where $A$ denotes the photon, $Z$ is the massive neutral gauge boson, and $Z'$ is the massless dark photon state. The relation between the gauge and mass basis is given by $V_i = \mathcal{O}_{ij} E_j$ such that 
$\mathcal{O}^T M_1^2 \mathcal{O} = {\rm Diag}(0, m_Z^2, 0)$ with \(m_Z^2\) given by Eq.~(\ref{eq:Zbosonmass}). The orthogonal transformation $\mathcal{O}$ can be explicitly constructed as
\be
\begin{aligned}
\mathcal{O} & = 
\begin{pmatrix} 
 c_1 & -s_1 & 0 \\
 s_1 &  c_1 & 0 \\
 0 & 0& 1
\end{pmatrix}
 \cdot
 \begin{pmatrix} 
 1 & 0 & 0 \\
 0 &  c_2 & -s_2 \\
 0 & s_2 &  c_2
\end{pmatrix} = 
\begin{pmatrix} 
 c_1 & -s_1 c_2 & s_1 s_2 \\
 s_1 & c_1 c_2  & -c_1 s_2  \\
 0 & s_2 & c_2
\end{pmatrix}, 
\end{aligned}
\label{eq:rotmatrix}
\ee
where
\bea
\label{eq:rotmatrixelements}
\begin{aligned}
s_1 \equiv \sin\theta_1 &=   \frac{\tilde{g}'}{g_M } \;, \;\;\;\; c_1 \equiv \cos\theta_1 = \frac{g}{g_M} \;, \\
s_2 \equiv \sin\theta_2 &= - \frac{\tilde{g}''}{\bar{g}_M} \;, \;\;\;\; c_2 \equiv \cos\theta_2 = \frac{g_M}{\bar{g}_M} \;.
\end{aligned}
\eea
where  $g_M = \sqrt{g^2 + \tilde{g}'^2}$ and $\bar{g}_M = \sqrt{g^2 + \bar{g}'^2}$. 
While $g_M$ depends on the rotation angle $\alpha$ through $\tilde{g}'$, $\bar{g}_M$ is independent of $\alpha$ because $\bar{g}'$ is. Diagonalization of the neutral gauge boson mass matrix results in modifications to the SM neutral current and introduce interactions between the dark photon and SM fermions. These neutral current interactions are flavor diagonal and given by
\be
\label{eq:LNC}
{\cal L}_{\rm NC} = \sum_i \sum_f \bar{f} \gamma_\mu \left( v_i^f - a_i^f \gamma_5 \right) f E_i^\mu \;,
\ee
where the vector and axial-vector couplings are given by
\bea
\label{eq:vi0}
v_i^f &=& \left( g \mathcal{O}_{2i} - \tilde{g}' \mathcal{O}_{1i}  - \tilde{g}'' \mathcal{O}_{3i} \right) \frac{T^3_f}{2} + (\tilde{g}'  \mathcal{O}_{1i} +  \tilde{g}''  \mathcal{O}_{3i})  Q_f \; , \\
\label{eq:ai0}
a_i^f &=& \left(  g \mathcal{O}_{2i} -  \tilde{g}' \mathcal{O}_{1i} - \tilde{g}''  \mathcal{O}_{3i}  \right) \frac{T^3_f}{2} \; ,
\eea
with $T^3_f$ denotes the quantum number of the left-handed chiral component of $f$ under $SU(2)_L$ and $Q_f = T^3_f + Y_f$ is the electric charge of \(f\).

Using the explicit expressions for the matrix elements of $\mathcal{O}$ in Eq.~(\ref{eq:rotmatrixelements}), one can obtain the vector and axial-vector couplings of the neutral gauge bosons to SM fermions as
\bea
\label{eq:neutralcoupling}
\begin{aligned}
v_{A}^f \equiv v_1^f &= \tilde{g}' c_1 Q_f = g s_1 Q_f \; , \\   
a_{A}^f \equiv  a_1^f &= 0 \; . \\
v_Z^f \equiv v_2^f &= \frac{g}{c_1 c_2} \left[ 
\frac{T^3_f}{2} - \left( s_1^2 c_2^2 + s_2^2 \right) Q_f \right] = \bar{g}_M \left( \frac{T^3_f}{2}  -  \frac{\bar{g}'^2}{\bar{g}_M^2} Q_f  \right) \; ,  \\
a_Z^f \equiv a_2^f &= \left( \frac{g}{c_1 c_2} \right)  \frac{T^3_f}{2} 
= \bar{g}_M\frac{T^3_f}{2}  \;  ,  \\
v_{Z'}^f \equiv v_3^f &= \tilde{g}'' c_1^2 c_2 Q_f = 
 \frac{\tilde{g}''  g^2}{g_M \bar{g}_M} Q_f \; , \\   
a_{Z'}^f \equiv  a_3^f &= 0 \; .
\end{aligned}
\eea
As expected, the axial-vector couplings of the photon and massless dark photon to SM fermions vanish. We also note that the $Z$-boson couplings to SM fermions, $v_Z^f$ and $a_Z^f$, are independent of the rotation angle $\alpha$, since they depend only on the $\alpha$-independent couplings $\bar{g}'$ and $\bar{g}'_M$. 
In contrast, the couplings $v_A^f$ and $v_{Z'}^f$ do depend on $\alpha$, but this $\alpha$ dependence cancels in physical observables.

The cubic and quartic self-interactions of all physical gauge bosons can be obtained from $\mathcal{L}_{\rm kin}$ in Eq.~(\ref{eq:kin}).
Recalling that the orthogonal transformation $\mathcal O$ between the gauge basis 
$V = ( B_\mu, W^3_\mu, W^3_{D\mu} )$ and the mass basis 
$E = ( A_\mu, Z_\mu, Z'_\mu )$ is given by
$V_{\mu i} = \mathcal{O}_{ij} E_{\mu j}$. Define the field strength as usual
$E_{\mu\nu} = (\partial_\mu E_{\nu} - \partial_\nu E_{\mu}) = (
F_{\mu\nu}, Z_{\mu\nu}, Z^\prime_{\mu\nu} )$.
The cubic term is given by
\bea
\begin{aligned}
\label{eq:L3g}
\mathcal{L}_{3g} & =  i g \mathcal{O}_{2i} \biggl[ 
\left( W^+_{\mu\nu} W^{- \mu} - W^-_{\mu\nu} W^{+\mu} \right) E_i^\nu 
+\frac{1}{2} \left( W^+_\mu W^-_\nu - W^-_\mu W^+_\nu \right) E_i^{\mu\nu} 
\biggr]  \\
& + 
i g_D \biggl[ \kappa_i
\left( W^p_{\mu\nu} W^{m \mu} - W^m_{\mu\nu} W^{p\mu} \right) E_i^\nu  +\frac{1}{2} \kappa'_i \left( W^p_\mu W^m_\nu -W^m_\mu W^p_\nu \right) E_i^{\mu\nu} 
\biggr] \; ,
\end{aligned}
\eea
where 
\bea
\label{eq:kappai}
\kappa_i &=& \frac{1}{\sqrt{1-\epsilon^2}} \left( -s_\alpha \mathcal{O}_{1i} + c_\alpha \mathcal{O}_{3i} \right) \; , \\
\label{eq:kappapi}
\kappa'_i &=&  
\left( \epsilon c_\alpha - \sqrt{1-\epsilon^2} s_\alpha \right)
\mathcal{O}_{1i} +  \left( \epsilon s_\alpha + \sqrt{1-\epsilon^2} c_\alpha \right) \mathcal{O}_{3i} \; .
\eea
From the first line of Eq.~(\ref{eq:L3g}), the overall coupling of the $Z W^+ W^-$ vertex is given by $g \mathcal{O}_{22} \equiv g c_1 c_2 = g^2/\bar g_M $, which is independent of $\alpha$. Similarly, the couplings associated with the $Z W^p W^m$ vertex in the second line of Eq.~(\ref{eq:L3g}),
\bea
\label{eq:kappa2}
\kappa_2&=& \frac{1}{\sqrt{1-\epsilon^2}} \left( -s_\alpha \mathcal{O}_{12} + c_\alpha \mathcal{O}_{32} \right) \equiv  \frac{\epsilon \bar{g}'}{\sqrt{1 - \epsilon^2} \bar{g}_M}\; , \\
\label{eq:kappap2}
\kappa'_2 &=&  
\left( \epsilon c_\alpha - \sqrt{1-\epsilon^2} s_\alpha \right)
\mathcal{O}_{12} +  \left( \epsilon s_\alpha + \sqrt{1-\epsilon^2} c_\alpha \right) \mathcal{O}_{32}  \equiv 0\; ,
\eea
are also independent of $\alpha$.

Lastly the quartic term of the gauge boson interactions can be written down as
\bea
\label{eq:L4g}
\begin{aligned}
\mathcal{L}_{4g} & =  - \frac{1}{2} g^2
\biggl[
\left( W^+_\mu W^{- \mu} \right)^2 - W^+_\mu W^{+ \mu} W^-_\nu W^{- \nu} \biggr. \\
& \;\;\;\; \biggl. + \, 2 \, \mathcal{O}_{2i} \mathcal{O}_{2j} \left( 
W^+_\mu W^{-\mu} E_{i\nu} E_j^\nu - E_{i\mu} W^{+\mu} E_{j \nu} W^{-\nu} \right) 
\biggr]  \\
& - \frac{1}{2} g_D^2
\biggl[
\left( W^p_\mu W^{m \mu} \right)^2 - W^p_\mu W^{p \mu} W^m_\nu W^{m \nu} \biggr. \\
& \;\;\;\; + \, 2 \, \kappa_{i}\kappa_j  
\left( 
W^p_\mu W^{m \mu} E_{i\nu} E_j^\nu - E_{i\mu} W^{p \mu} E_{j \nu} W^{m \nu} \right) 
\biggr] \; .
\end{aligned}
\eea
Similar to the cubic terms, the quartic vertices $Z Z W^+ W^-$ and $Z Z W^p W^m$ are also independent of $\alpha$.

\section{Rotation angle independence and millicharged basis}
\label{sec:milichargedbasis}

In the previous section, we showed that all interactions involving the massive neutral gauge boson $Z$ are independent of the rotation angle $\alpha$. However, this property is not manifest for the two massless states $\gamma$ and $Z'$. In this section, we demonstrate that physical observables involving $\gamma$ and $Z'$ are also independent of $\alpha$.

As discussed in Ref.~\cite{Pan:2018dmu} for Abelian kinetic-mixing with a massless dark photon, the identification of the physical photon and dark photon is basis dependent, as reflected by the dependence on the rotation angle $\alpha$. Consequently, individual amplitudes generally depend on $\alpha$. However, physical observables must include all contributions from the degenerate massless gauge-boson sector. After summing over all physically equivalent initial/final or intermediate states in the squared amplitudes, all observables become independent of the basis choice, {\it i.e.} the $\alpha$-dependence cancels.
We observe the same feature in the present Abelian--non-Abelian kinetic-mixing scenario with a massless dark photon.

For example, the process $W^+W^- \to$ two real photons receives contributions from the $t$- and $u$-channel $W^\pm$ exchange diagrams and the seagull diagram for each of the three final states $\gamma\gamma$, $\gamma Z'$, and $Z'Z'$, such that
\be
\label{eq:ww2photons}
\begin{aligned}
\sum_{\rm pol.} |M|^2_{W^+ W^- \to 2\, {\rm photons} } 
& =
\frac{1}{2} \left(  \sum_{\rm pol.} |M|^2_{W^+ W^- \to \gamma \gamma}
+
\sum_{\rm pol.} |M|^2_{W^+ W^- \to Z' Z'} \right) \\
& \qquad +
\sum_{\rm pol.} |M|^2_{W^+ W^- \to \gamma Z'} 
\end{aligned}
\ee
where the factors of $1/2$ account for identical particles in the final state.
Extracting the overall gauge-coupling dependence, one finds
\be
\label{eq:ww2photons2}
\begin{aligned}
\sum_{\rm pol.} |M|^2_{W^+ W^- \to 2\, {\rm photons} }
& =
g^4
\left(
{\cal O}_{21}^4
+
{\cal O}_{23}^4
+
2 {\cal O}_{21}^2 {\cal O}_{23}^2
\right) \times F \\
& =
g^4
\left(
1-c_1^2 c_2^2
\right)^2 \times F \\
& =
\frac{g^4 \bar g'^4}{\bar g_M^4} \times F \\
& =
e^4 \times F 
\end{aligned}
\ee
where \(F\) is the common invariant function of the Mandelstam variables \(s,t,\) and \(u\), and
\begin{equation}
\label{eq:edefspecial}
e = g \left( \frac{\bar g'}{\bar g_M} \right)
\end{equation}
defines the electric charge \(e\), which is independent of the rotation angle $\alpha$. 

Similarly, one may consider the process \(W^p W^m \to \) two real photons and obtain
\be
\label{eq:WpWm2photons}
\begin{aligned}
\sum_{\rm pol.} |M|^2_{W^p W^m \to 2\, {\rm photons} } 
& = \frac{1}{2} \left( \sum_{\rm pol.} |M|^2_{W^p W^m \to \gamma \gamma}
+ \sum_{\rm pol.} |M|^2_{W^p W^m \to Z' Z'} \right) \\
& \qquad  + \sum_{\rm pol.} |M|^2_{W^p W^m \to \gamma Z'}  
 \\
& = 
g_D^4 \Big[ \left( \kappa_1^2 + \kappa_3^2 \right)^2 F_1 
+ \left( \kappa^{\prime 2}_1 + \kappa^{\prime 2}_3 \right)^2 F_2
 \\
& \qquad +  \left( \kappa_1^2 + \kappa_3^2 \right) \left( \kappa^{\prime 2}_1 + \kappa^{\prime 2}_3 \right) F_3 
+ \left(  \kappa_1 \kappa'_1 + \kappa_3 \kappa'_3 \right)^2 F_4
 \\
& \qquad +  \left( \kappa_1^2 + \kappa_3^2 \right)  \left(  \kappa_1 \kappa'_1 + \kappa_3 \kappa'_3 \right) F_5 
\\
 & \qquad + \left( \kappa^{\prime 2}_1 + \kappa^{\prime 2}_3 \right)  \left(  \kappa_1 \kappa'_1 + \kappa_3 \kappa'_3 \right) F_6 
\Big]
\end{aligned}
\ee
where $F_i \, (i=1,\cdots,6)$ are invariant functions of the Mandelstam variables $s$, $t$, and $u$. 
Using Eqs.~\eqref{eq:kappai} and~\eqref{eq:kappapi}, one finds 
\bea
\begin{aligned}
\kappa_1^2 + \kappa_3^2 &= 1 + \frac{\epsilon^2 g^2}{(1 - \epsilon^2) (g^2 + \bar{g}'^2)} = \frac{g^2 + g'^2}{g^2(1-\epsilon^2) + g'^2} \; , \\
\kappa^{\prime 2}_1 + \kappa^{\prime 2}_3  &= 1 \; ,  \\
\kappa_1 \kappa'_1 + \kappa_3 \kappa'_3   &= 1 \; ,
\end{aligned}
\eea
which explicitly shows that the process \(W^p W^m \to \) two real photons is also independent of $\alpha$.

Therefore, although the individual amplitudes depend on the basis choice, the corresponding physical observables are independent of $\alpha$. We conjecture that the same conclusion holds for arbitrary tree-level processes involving degenerate massless photon and dark-photon states in the initial/final or intermediate states, provided that all relevant contributions in the squared amplitudes are properly summed.

Since the parameter $\alpha$ corresponds to a redundant rotation in gauge space, one may choose a convenient basis to simplify the phenomenological analysis. Hereafter, we adopt the millicharged basis defined by
\begin{equation}
\label{eq:millichargedbasis}
s_\alpha = \epsilon \, ,
\qquad
c_\alpha = \sqrt{1-\epsilon^2} \, ,
\end{equation}
for which
\begin{equation}
\tilde g' = \bar g' = \frac{g'}{\sqrt{1-\epsilon^2}} \, ,
\qquad
\tilde g'' = 0 \; .
\end{equation}
In this ``millicharged basis'', the two rotation angles in the matrix \(\mathcal{O}\) simplify to 
\begin{equation}
\begin{aligned}
s_1 & = \frac{\bar g'}{\sqrt{g^2 + \bar g'^2}} \, , & c_1 = \frac{g}{\sqrt{g^2 + \bar g'^2}} \; , \\
s_2 & = 0 \, , & c_2 = 1   \; . 
\end{aligned}
\end{equation}
Furthermore, Eq.~\eqref{eq:neutralcoupling} shows that the massless dark photon does not couple to SM fermions at tree level~\footnote{
Effective interactions between the massless dark photon and ordinary matter can still be induced through higher-dimensional operators~\cite{Dobrescu:2004wz}.
},
while the photon couplings retain exactly the same form as in the SM with \(e = g s_1\)~\footnote{The electric charge takes the same form as in the SM, with the replacement $g' \to \bar{g}' = g'/\sqrt{1 - \epsilon^2}$.
}. One can readily verify that the general definition of \(e\) given in Eq.~(\ref{eq:edefspecial}) also reduces to the SM form.

Similarly, from the cubic gauge interaction in Eq.~\eqref{eq:L3g}, the $\gamma W^+W^-$ coupling is given by
\begin{equation}
g {\cal O}_{21} \equiv g s_1 = e \, ,
\end{equation}
while the quartic $\gamma\gamma W^+W^-$ interaction from Eq.~\eqref{eq:L4g} is proportional to
\begin{equation}
g^2 {\cal O}_{21}^2 = e^2 \, .
\end{equation}
Consequently, both the cubic $\gamma W^+W^-$ and quartic $\gamma\gamma W^+W^-$ vertices preserve the same structure as in the SM. On the other hand, the cubic $Z'W^+W^-$ and quartic $Z'Z'W^+W^-$ interaction vertices vanish identically.

The couplings between the complex dark gauge bosons $W^{(p,m)}$ and the neutral massless gauge bosons also simplify considerably. From Eqs.~\eqref{eq:kappai} and~\eqref{eq:kappapi}, one obtains
\bea
\label{eq:kappa123}
\begin{aligned}
\kappa_1 &= - \frac{\epsilon c_1}{\sqrt{1-\epsilon^2}} \; ,
\qquad &
\kappa'_1 = 0 \; , \\
\kappa_2 & =  + \frac{\epsilon s_1}{\sqrt{1-\epsilon^2}} \; ,
\qquad &
\kappa'_2 = 0  \; , \\
\kappa_3 &= 1 \; ,
\qquad &
\kappa'_3 = 1 \; .
\end{aligned}
\eea
Therefore, the interactions between $W^{(p,m)}$ and the photon are suppressed by the kinetic-mixing parameter $\epsilon$, leading to an effective millicharge interaction for the dark charged vector bosons.

The millicharged dark vectors $W' = W^{p,m}$ are stable due to the unbroken $U(1)_D$ charge, making them natural dark matter candidates based on gauge symmetry. Their self-interactions are dominantly mediated by the long-range massless dark photon and are controlled by the hidden gauge coupling $g_D$.

\section{Freeze-in Production}
\label{sec:FreezeIn}

Because the portal couplings $\epsilon, g_D \ll 1$, the hidden sector never thermalizes with the SM bath. Instead, dark particles are produced gradually via freeze-in, sourced by SM annihilations, decays, and in-medium processes such as plasmon decay. As matter accumulates in the hidden sector, interactions among hidden particles become relevant and proceed concurrently with freeze-in. If kinematically allowed, hidden states may also decay back into SM particles.

We employ a two-temperature Boltzmann framework, with $T$ and $T_h$ denoting the visible and hidden temperatures. The evolution of the ratio $\xi = T_h/T$, accounting for feeble energy transfer, is governed by~\cite{Aboubrahim:2020lnr,Aboubrahim:2021ycj,Aboubrahim:2021dei,Li:2023nez} (see Appendix~\ref{app:dxidT} for detailed derivations.)
\begin{equation}
\label{eq:dxidT}
\frac{d\xi}{dT} = \frac{1}{T}\left[ -\xi + \left(\frac{3H(\rho_h + p_h) - J_h}{3H(\rho_v + p_v) + J_h}\right) \frac{d\rho_v}{dT} \left(\frac{d\rho_h}{dT_h} \right)^{-1} \right] \; ,
\end{equation}
where $H$ is the Hubble parameter, and $\rho_{v,h}$ and $p_{v,h}$ are the energy and pressure densities of the visible and hidden sectors (see Appendix~\ref{app:rho_p_s} for explicit forms). The source function $J_h$ encodes all energy-exchange processes between the two sectors; its explicit form is given in Appendix~\ref{app:Jh}.

The relic abundance is obtained by solving coupled Boltzmann equations for the comoving yields $Y_i=n_i/\mathbf{s}$,
\begin{equation}
\label{eq:dYidT}
\frac{dY_{i}}{d T} = -\frac{1}{\mathbf{s}} \left[\frac{d\rho_v/dT}{3H(\rho_v + p_v) + J_h} \right] \, {\mathcal{R}}_{i}(T,T_h) \; ,
\end{equation}
where $\mathbf{s}$ is the total entropy density, $n_i$ the number density, and $\mathcal{R}_i(T,T_h)$ the interaction rate producing species $i$, including all three SM$\to$hidden, hidden$\to$SM, and hidden$\to$hidden channels. We track all three species: the dark matter candidate with $n_{W^m}=n_{W^p}\equiv n_{W^\prime}$, the dark photon, and the hidden Higgs. Explicit expressions for the rates $\mathcal{R}_i$, including cross sections and decay widths, are provided in Appendices~\ref{app:CoupledBoltzmann},~\ref{app:xsec_decay} and~\ref{app:plasmondecay}. Plasmon decay (Appendix~\ref{app:plasmondecay}) is included but found to be subdominant.

\section{Galaxy Dynamics, Cosmological Constraints, and Direct Detection}
\label{sec:Constraints}

A massless dark photon mediates long-range DM self-interactions that are strongly constrained by galactic dynamics. 
Excessively strong dark matter self-interactions would erase the observed anisotropy of galaxies.
The observed ellipticity of halos, {\it e.g.}~NGC720, requires~\cite{Agrawal:2016quu,Fabbrichesi:2020wbt}
\be
m_{\rm DM}\!\left(\tfrac{0.01}{\alpha_D}\right)^{2/3} \gtrsim 300~{\rm GeV} \; , \;\;\; \alpha_D = \frac{g_D^2}{4 \pi}
\label{eq:ellipticity}
\ee
ensuring that halo shapes remain intact.

Cosmological observations further constrain the hidden sector. In particular, the number of additional relativistic degrees of freedom—such as a massless dark photon or a Goldstone boson—is tightly constrained by measurements from the Planck Collaboration~\cite{Planck:2019nip,Planck:2018vyg}, since an excess of such species would increase the Hubble expansion rate. A faster expansion reduces the neutron-proton conversion rate, leading to a larger neutron abundance at freeze-out. As a result, the primordial helium abundance, which is approximately twice the neutron abundance, would be enhanced and thus become inconsistent with observations.
Light states contribute to the effective number of relativistic degrees of freedom~\cite{Li:2023nez},
\be
\Delta N_{\rm eff}(T,\xi) = \tfrac{4}{7}\, g_{\rm eff}^h(T) \left(\tfrac{11}{4}\right)^{4/3}\xi^4 \;,
\label{eq:Neff}
\ee
with $\xi=T_h/T$. Current BBN and CMB analyses demand $\Delta N_{\rm eff} < 0.18$ at 95\% CL~\cite{Yeh:2022heq}. In addition, the dark Higgs must decay before nucleosynthesis to avoid disrupting light-element abundances, typically requiring $\tau_{h_D}\lesssim 0.1$\,s for $2m_\mu \lesssim m_{h_D} \lesssim m_h/2$~\cite{Fradette:2017sdd}.

Finally, sub-GeV millicharged DM is subject to direct-detection searches targeting electron recoils. The leading limits arise from SENSEI~\cite{SENSEI:2020dpa}, DAMIC~\cite{DAMIC-M:2023gxo}, EDELWEISS~\cite{EDELWEISS:2020fxc}, SuperCDMS~\cite{SuperCDMS:2018mne}, CDEX~\cite{CDEX:2022kcd}, DarkSide-50~\cite{DarkSide:2022knj}, XENONnT~\cite{XENON:2024znc}, and PandaX-II~\cite{PandaX-II:2021nsg}, while OSCURA~\cite{Oscura:2022vmi,Oscura:2023qik} and novel superconducting or polar-material detectors~\cite{Hochberg:2021pkt,Knapen:2021run,Knapen:2017ekk,Griffin:2018bjn} will extend coverage. Defining $ \mu_{W' e} = m_{W'}m_e/(m_{W'}+m_e)$ and $\alpha_{\rm EM} = e^2/4\pi$, the constraints are conveniently expressed in terms of the reference elastic cross section $W' e^- \to W' e^-$,
\be
\overline{\sigma}_e = \frac{4 g_D^2 \kappa_1^2 \,\alpha_{\rm EM}\, \mu_{W^\prime e}^2}{(\alpha_{\rm EM} m_e)^4} \; , 
\ee
which is dominated by photon exchange, with $Z$- and scalar-mediated contributions being subleading.

\begin{figure*}[htbp!]
    \centering
    \includegraphics[width=0.95\textwidth]{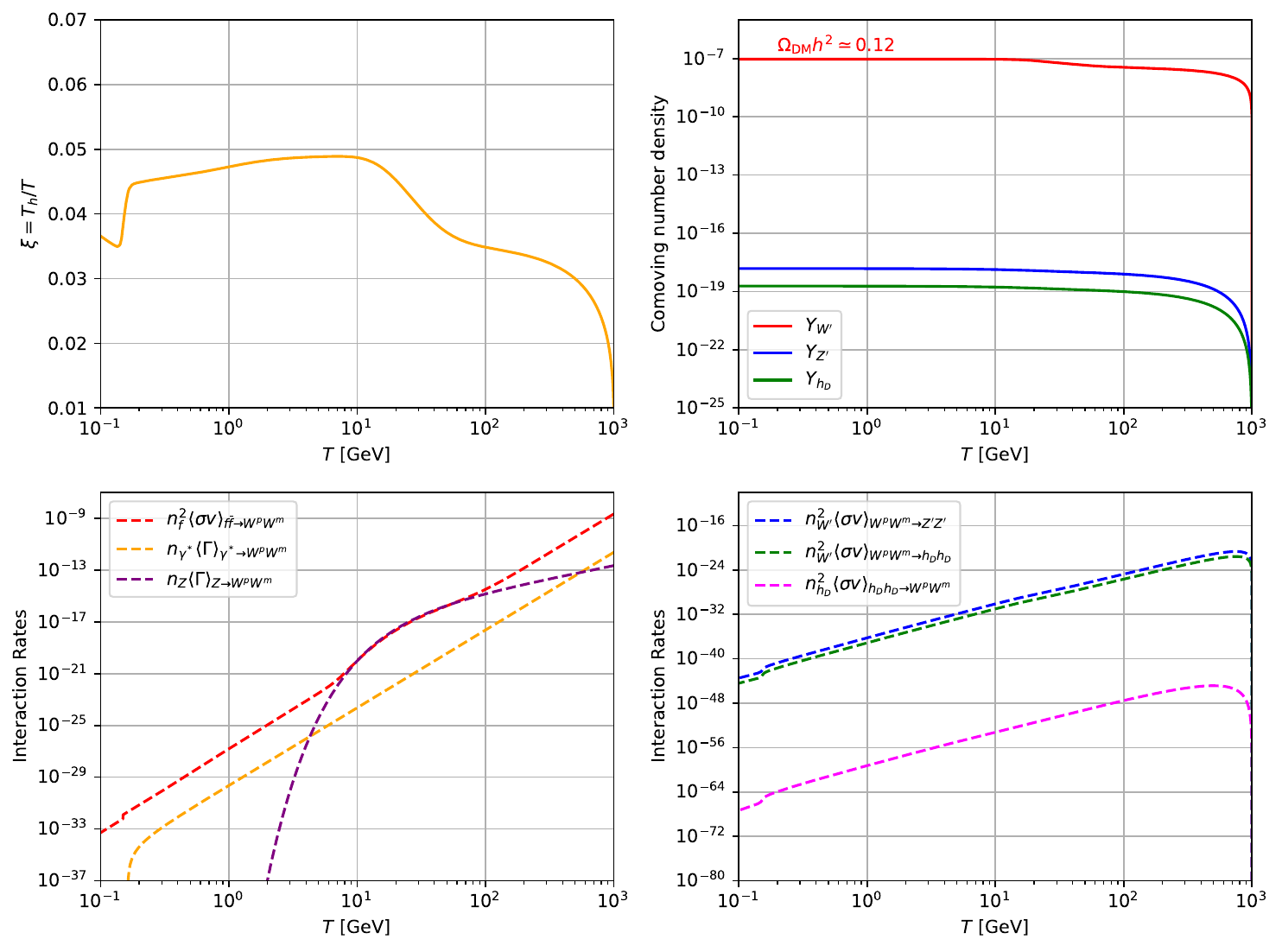}
    \caption{ \label{fig:BM1evolution} 
    Evolution of the dark sector for benchmark point BM1, defined by $m_{Z^\prime} = 0~\text{GeV}$, $m_{W^\prime} = 0.01~\text{GeV}$, $m_{h_D} = m_{W^\prime}/3$, $g_D = 10^{-7}, \epsilon = 3.8 \times 10^{-7}$, and $\sin\beta = 0$. Top left: temperature ratio $\xi = T_h/T$ between the dark and visible sectors. Top right: comoving number densities of dark matter (solid red), dark photons (solid blue), and dark Higgs bosons (solid green). Bottom panels: interaction rates (per unit volume per unit time) for various processes: 
    $f\bar{f} \to W^p W^m$ (red dashed), $Z$-boson decay (purple dashed), plasmon decay (orange dashed), dark matter pair annihilation into $Z^\prime Z^\prime$ (blue dashed) and $h_D h_D$ (green dashed), and the inverse process $h_D h_D \to W^p W^m$ (magenta dashed).
    }
\end{figure*}

\begin{figure*}[!htbp]
    \centering
    \includegraphics[width=0.95\textwidth]{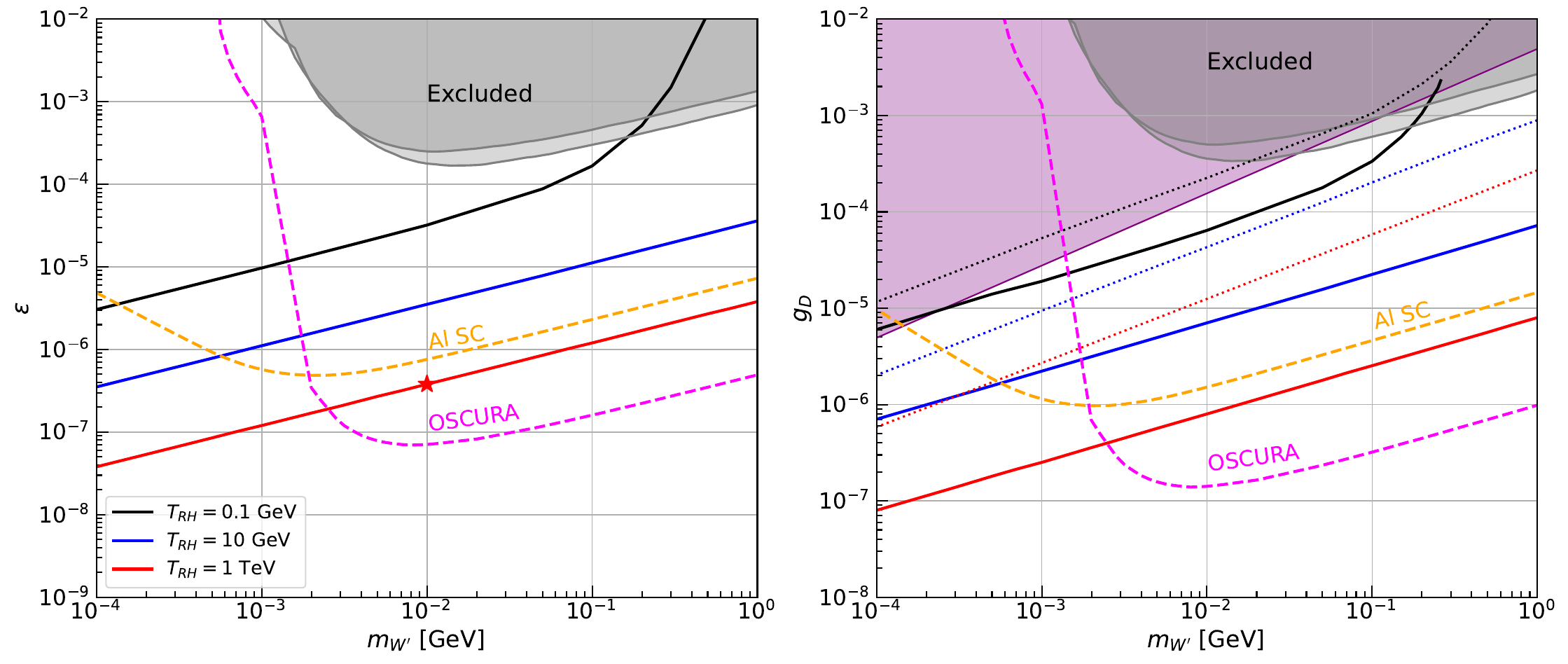}
    \caption{ \label{fig:mWp_eps_BM1} 
    Parameter space yielding the correct freeze-in dark matter abundance with reheating temperatures $T_{\rm RH} = 0.1~\text{GeV}, 10~\text{GeV} $, and $ 1~\text{TeV}$ represented by solid black, blue, and red curves, respectively. 
    {Left:} projection onto the $(m_{W^\prime},\epsilon)$ plane with $g_D = 10^{-7}$.  
    {Right:} projection onto $(m_{W^\prime}, g_D)$ plane with $\epsilon = 0.5\times10^{-7}$.  
    All remaining parameters follow benchmark {BM1}, and we fix $\xi_0 = 0.01$.
    The purple shaded region is excluded by halo-shape (ellipticity) constraints described in~(\ref{eq:ellipticity}).
 The gray region is excluded by current DM direct detection searches from SENSEI~\cite{SENSEI:2020dpa}, DAMIC~\cite{DAMIC-M:2023gxo}, EDELWEISS~\cite{EDELWEISS:2020fxc}, SuperCDMS~\cite{SuperCDMS:2018mne}, CDEX~\cite{CDEX:2022kcd}, DarkSide-50~\cite{DarkSide:2022knj}, XENONnT~\cite{XENON:2024znc}, and PandaX-II~\cite{PandaX-II:2021nsg}.  
    The dashed magenta and orange lines show the projected reach of OSCURA~\cite{Oscura:2022vmi, Oscura:2023qik} and aluminum-based superconducting detectors~\cite{Hochberg:2021pkt, Knapen:2021run}, respectively.
 The dotted black, blue, and red lines indicate the parameter space where dark-sector annihilations significantly deplete the freeze-in abundance with reheating temperatures \( T_{\rm RH} = 0.1~\text{GeV}, 10~\text{GeV} \), and \( 1~\text{TeV} \), respectively. The red star point in the left panel represents the benchmark {BM1}.
    }
\end{figure*}

\section{Numerical Results}
\label{sec:NumericalResults}

We now present numerical results for freeze-in production of the vector dark matter \(W^{p,m}\) in the Abelian--non-Abelian kinetic-mixing model. 
We choose to work in the ``millicharged basis'' defined by Eq.~(\ref{eq:millichargedbasis}) so that the massless dark photon does not couple to SM fermions.
As our first representative case, we define benchmark point BM1 with 
\(
m_{W^\prime} = 0.01~\text{GeV},\ m_{h_D} = m_{W^\prime}/3,\ g_D = 10^{-7},\ \epsilon = 3.8\times10^{-7},\ \sin\beta = 0 \,.
\)  
In this scenario, the dark sector communicates with the SM solely via millicharged interactions,
with $g_D$ constrained by halo-shape bounds and $\epsilon$ adjusted to reproduce the observed relic density, taking $T_0 = T_{\rm RH} = 1$~TeV and an initial temperature ratio $\xi_0 = (T_h/T)_0 = 0.01$.

Fig.~\ref{fig:BM1evolution} shows the thermal history for BM1. The hidden sector starts empty and gradually populates as the Universe cools. Production is dominated by millicharged annihilation $f\bar f \to W^p W^m$ (red dashed), while $Z$-boson and plasmon decays provide subleading contributions that switch off once the plasma cools. A mild enhancement at $T \sim 10$–$100$~GeV arises from $Z$ exchange, and the kink at $T\simeq150$~MeV reflects the QCD transition. DM annihilation into dark photons or the hidden Higgs is suppressed by the small $g_D$, leaving $W^{p,m}$ as the dominant relic component. As the rates fall below the Hubble expansion, all abundances freeze in, reproducing the observed density $\Omega_{\rm DM} h^2 = 0.120\pm0.001$~\cite{Planck:2018vyg}.

We then explore deviations from {BM1} by varying the reheating temperature, initial temperature ratio \(\xi_0\), and other model parameters. 
 The left panel of Fig.~\ref{fig:mWp_eps_BM1} shows the viable region in the $(m_{W^\prime},\epsilon)$ plane, keeping other parameters as in {BM1}. Since the production rate from the dimension-5 operator scales as ${\cal R}_{W^\prime} \sim T^6/\Lambda^2$, freeze-in remains efficient up to $T_{\rm RH}$, making the relic abundance sensitive to $T_{\rm RH}$: higher reheating temperatures increase production, requiring larger $\Lambda$ (or smaller $\epsilon$) to match observations. For $m_{W^\prime} \gtrsim T_{\rm RH}$, production is Boltzmann suppressed, and only very large $\epsilon$ could compensate—values already excluded by collider, neutrino, and beam-dump searches~\cite{OPAL:1995uwx,Prinz:1998ua,ArgoNeuT:2019ckq,Ball:2020dnx,Magill:2018tbb,Marocco:2020dqu}. Current direct-detection limits already constrain part of the low-$T_{\rm RH}$ region (gray area in Fig.~\ref{fig:mWp_eps_BM1}), while next-generation sub-GeV detectors such as OSCURA (magenta dashed) and aluminum-based superconducting targets (orange dashed) will probe much of the freeze-in parameter space.

 The right panel of Fig.~\ref{fig:mWp_eps_BM1} displays the viable region in the $(m_{W'},g_D)$ plane for fixing $\epsilon = 0.5\times10^{-7}$ and keeping other parameters as in {BM1}.
Large gauge couplings are restricted by dark-matter self-interaction limits (purple), while the gray region is ruled out by current sub-GeV direct-detection experiments.
Crucially, the dotted curves denote the boundary of dark-sector thermalization for different reheating temperatures.
Above these lines, processes such as $W'W' \to Z'Z'$ and $W'W' \to h_D h_D$ efficiently erase the freeze-in population.

\begin{figure}[!htbp]
    \centering
    \includegraphics[width=0.65\textwidth]{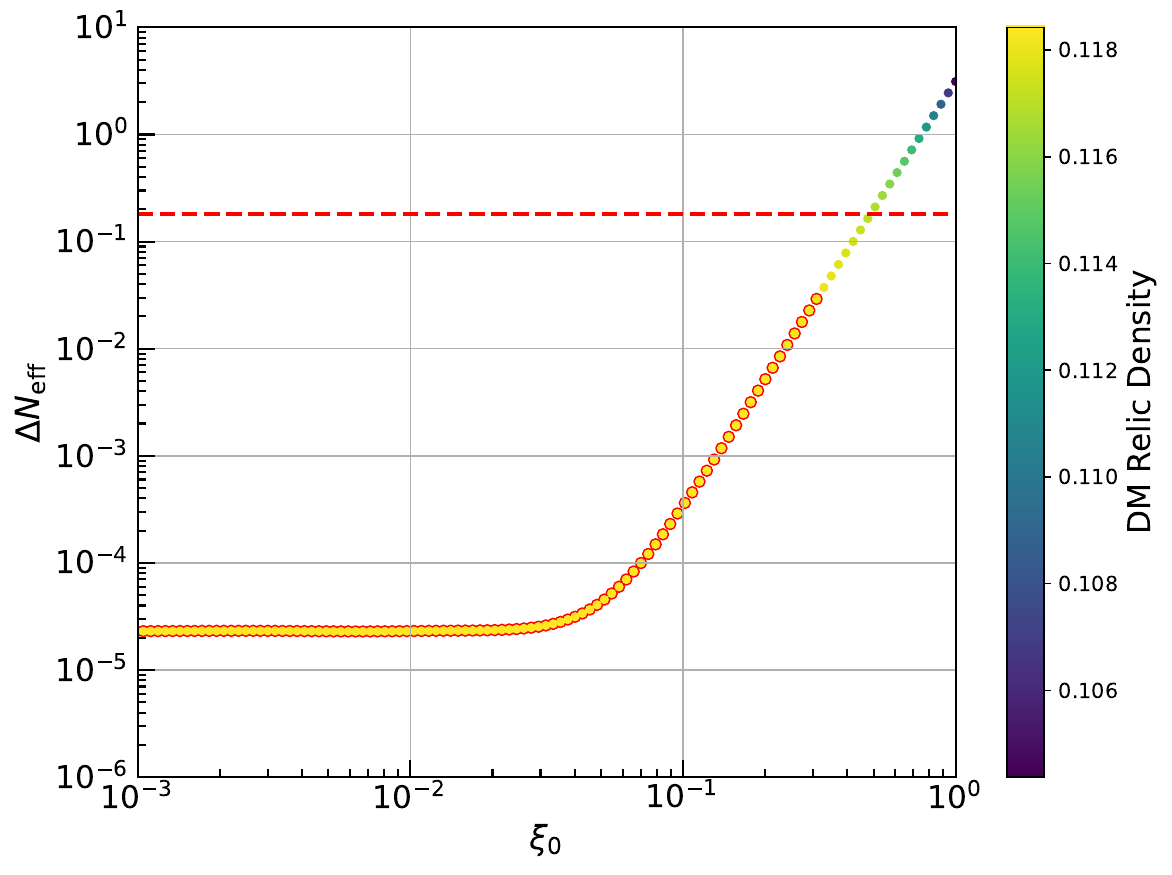}
    \caption{ \label{fig:DNeff_BM1} 
    Effective number of relativistic species, \( \Delta N_{\rm eff}  (T_{\rm BBN}) \) evaluated at the BBN, and dark matter relic density (indicated by the color scale) as functions of the initial temperature ratio $\xi_0 = (T_{h0}/T_0)_{\rm RH}$ with $T_0 = T_{\rm RH} =1$ TeV. All other parameters are fixed to the values of benchmark point {BM1}. The region above the red dashed line corresponds to $\Delta N_{\rm eff} > 0.18$, which is excluded by combined BBN and CMB constraints~\cite{Yeh:2022heq}. Points highlighted with red circles indicate parameter choices that reproduce the observed dark matter relic density within the $2\sigma$ range.
    }
\end{figure}

The impact of the initial hidden-to-visible temperature ratio is illustrated in Fig.~\ref{fig:DNeff_BM1}. Larger $\xi_0$ enhances the hidden radiation contribution to  $\Delta N_{\rm eff}(T_{\rm BBN})$ while suppressing the DM relic abundance. Values $\xi_0 \gtrsim 0.5$ are excluded by BBN and CMB bounds ($\Delta N_{\rm eff} < 0.18$ at 95\% C.L.), whereas $\xi_0 \lesssim 0.03$ gives a negligible contribution below current sensitivity. Cosmology thus tightly constrains the initial hidden-sector conditions while leaving ample parameter space consistent with freeze-in.

\begin{figure}[htbp!]
    \centering
    \includegraphics[width=0.95\textwidth]{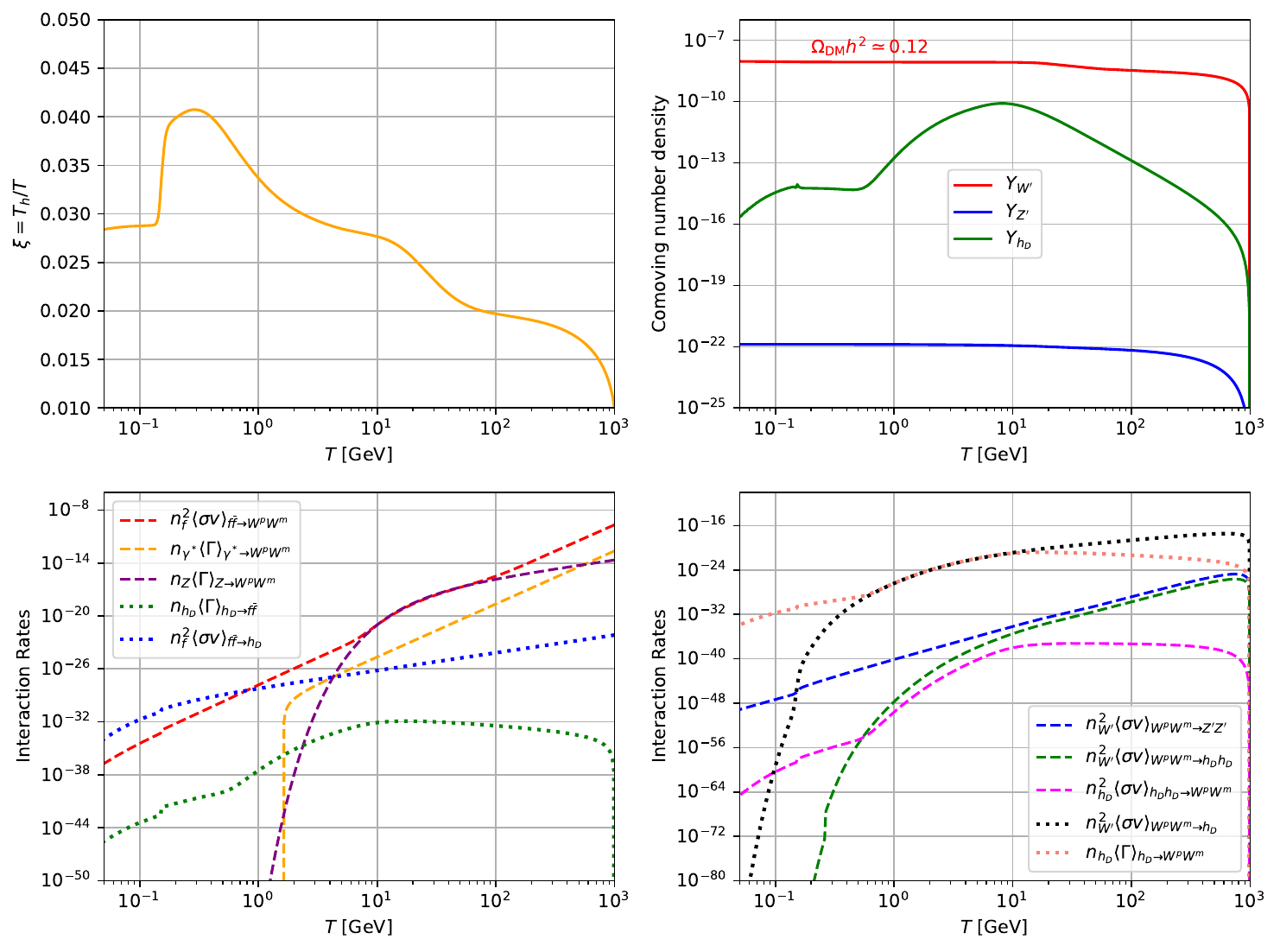}
    \caption{ \label{fig:BM2evolution} 
    Evolution of the dark sector for benchmark point {BM2}, defined by $m_{Z'} = 0~\text{GeV}, m_{W'} = 0.1~\text{GeV}, m_{h_D} = 0.4~\text{GeV}, g_D = 10^{-7}, \epsilon = 1.2 \times 10^{-6}$,  and $\sin\beta = 10^{-9}$. {Top left}: temperature ratio $\xi = T_h/T$ between the dark and visible sectors. {Top right}: comoving number densities of dark matter (solid red), dark photons (solid blue), and dark Higgs bosons (solid green). {Bottom panels}: interaction and decay rates (per unit volume per unit time) for the following processes: $f\bar{f} \to W^pW^m$ (red dashed), $Z$-boson decay (purple dashed), plasmon decay (orange dashed), dark matter pair annihilation into $Z'Z'$ (blue dashed) and $h_D h_D$ (green dashed), inverse annihilation $h_D h_D \to W^pW^m$ (magenta dashed), dark Higgs production from DM (black dotted) and SM fermion (blue dotted) annihilation, and dark Higgs decays into DM (orange dotted) and SM fermions (green dotted).
    }
\end{figure}

Next, we examine a second benchmark point, {BM2}, defined by the parameter set $m_{Z'} = 0$ GeV, $m_{W^\prime} = 0.1~\text{GeV}$, $m_{h_D} = 0.4~\text{GeV}$, $g_D = 10^{-7}$, $\epsilon = 1.2 \times 10^{-6}$, and $\sin\beta = 10^{-9}$. The initial temperature is set to $T_0 = 1~\text{TeV}$, with an initial temperature ratio $\xi_0 = 0.01$, identical to the setup in~{BM1} shown in the main text. The evolution of the dark sector for this benchmark is shown in Fig.~\ref{fig:BM2evolution}. Compared to {BM1}, the presence of nonzero Higgs dark scalar mixing introduces an additional portal interaction between the dark sector and the SM. Moreover, since the dark Higgs mass exceeds twice the dark matter mass, it can decay into a pair of vector DM, while the reverse process DM annihilation into a dark Higgs is also kinematically allowed. 
The effects of this additional portal in { BM2} are clearly reflected by the richer features in the evolution of \( \xi \) and the enhancement of dark Higgs boson yield in the top left and top right panels respectively in Fig.~\ref{fig:BM2evolution}, as compared with {BM1}.

As shown in the bottom left panel of Fig.~\ref{fig:BM2evolution}, the dark Higgs can now be produced via annihilation of SM fermions, with interaction rates depicted by the blue dotted line. Additionally, as the DM number density builds up from the millicharged freeze-in process, DM particles begin to annihilate into dark Higgs bosons, as indicated by the black dotted line in the bottom right panel. This results in an increase in the dark Higgs number density just after the reheating temperature, as shown by the green line in the top right panel.

However, this increase halts around $T \sim 10$ GeV, where the decay rate of the dark Higgs into DM exceeds the Hubble rate, leading to a depletion of the dark Higgs population. As the Universe continues to cool, dark Higgs production becomes increasingly suppressed, while its decay remains efficient, resulting in a continued decrease in its number density. Notably, although the dark Higgs can decay into SM fermion pairs (green dotted line, bottom left panel), this channel is negligible compared to its dominant decay into DM (orange dotted line, bottom right panel).
In {BM2}, the lifetime of the dark Higgs is $\tau_{h_D} = 7.49 \times 10^{-10}$s, which satisfies the BBN constraints discussed in the main text.

As in {BM1}, the dark photon number density remains small due to the tiny gauge coupling $g_D$, which suppresses DM annihilation. Nevertheless, the observed DM relic abundance (red line, top right panel) is successfully reproduced through the dominant freeze-in process $f\bar{f} \to W^p W^m$ (red dashed line, bottom left panel).

\section{Conclusion}
\label{sec:Conclusions}

Hidden-sector millicharged vector gauge particles constitute an intricate yet intriguing dark matter possibility, particularly given that all known elementary charged particles carry charges quantized in units of $e$ or $e/3$.

In this work, we have presented a minimal freeze-in scenario in which dark matter consists of a massive dark vector particle-antiparticle pair arising from a hidden non-Abelian $SU(2)$ gauge sector spontaneously broken to a residual \(U(1)\) symmetry with a massless dark photon mediator. The pair acquires tiny millicharges through a dimension-4 kinetic-mixing term with an effective coefficient \( \epsilon \), generated by a higher-dimensional dimension-5 operator.

Owing to the degeneracy between the massless dark photon and the SM photon, the arbitrary rotation angle \( \alpha \), permitted in the general linear transformation that diagonalizes the kinetic-mixing term, enters various interaction vertices as well as individual amplitudes involving the two degenerate states. We showed that this $\alpha$-dependence cancels once the equivalent squared amplitudes are summed appropriately in our Abelian–non-Abelian kinetic-mixing setup, thereby generalizing previous results in the literature obtained for purely Abelian scenarios.

The hidden gauge sector interacts feebly with the SM through the kinetic mixing \( \epsilon \), enabling freeze-in production from the SM plasma, while its internal interaction strength is set by \( g_D \). For \(\epsilon, g_D \sim 10^{-7}\), the observed relic abundance can be reproduced through a two-temperature Boltzmann evolution while remaining consistent with astrophysical and cosmological constraints. Plasmon decay into the invisible vector particle-antiparticle dark matter pair were also included in our analysis, although their contribution was found to be subdominant. Interestingly, the viable freeze-in parameter space of this model lies within the projected sensitivity of upcoming sub-GeV direct-detection experiments.

In conclusion, this framework provides a predictive and experimentally testable scenario linking hidden non-Abelian dynamics, long-range dark interactions, and vector freeze-in dark matter.

\section*{Acknowledgments}
V.Q.T. and T.C.Y. are supported in part by the National Science and Technology Council (NSTC) of Taiwan under Grants No.~112-2811-M-001-089 and No.~113-2112-M-001-001, respectively, with additional support for V.Q.T. from the Ministry of Education through the Higher Education Sprout Project (No.~NTU-114L104022-1) and the National Center for Theoretical Sciences (NCTS).
T.C.Y. gratefully acknowledges the generous hospitality of Professor Nguyen Nhu Le at Hue University of Education, Hue, Vietnam, and Professor Khiem Hong Phan at Duy Tan University, Danang, Vietnam, where substantial progress on this work was made.

\appendix
\allowdisplaybreaks

\section{UV Completion for the Dimension-5 Operator}
\label{app:UVcompletion}

A minimal ultraviolet (UV) completion that generates the dimension-5 operator
in the main text can be obtained by introducing a vectorlike Dirac lepton
$\Psi(\mathbf{1},\mathbf{2},y)$,
which is an \(SU(2)_D\) doublet carrying hypercharge \(y\)
and is a singlet under \(SU(2)_L\).
Its mass and interactions are described by
\be
\label{eq:Lag_Psi}
{\cal L}_{\Psi}
=
\bar\Psi\,(i\slashed{D}-M_\Psi)\Psi
-\!\left(
y_{\Psi}\,\bar\Psi\,\Sigma\,\Psi
+ \mathrm{H.c.}
\right) \; ,
\ee
where the covariant derivative is
\be
D_\mu \Psi = (\partial_\mu 
  - i g_D W_{D\mu}^a T_D^a 
  - i g' y B_\mu ) \Psi \; , 
 \ee
 \(y_{\Psi} \) is a complex Yukawa coupling,
 \(M_\Psi\) is heavy lepton mass
and \(\Sigma = \Sigma^a T_D^a\).
The Yukawa terms in Eq.~(\ref{eq:Lag_Psi})
splits the heavy lepton doublet mass once
\(\Sigma\) acquires a VEV.

Integrating out the heavy lepton at one loop
induces a gauge-invariant
dimension-5 kinetic-mixing operator
connecting the non-Abelian and Abelian field strengths.
The relevant triangle diagram contains
one \(\Sigma\) insertion and two external gauge bosons
\(W_D^{\mu\nu}\) and \(B^{\mu\nu}\).
In the low-energy limit \(|p^2| \ll M_\Psi^2\),
matching the amplitude to dimension-5 operator yields~\cite{Ardu:2024bxg}
\begin{equation}
 \frac{c_5}{\Lambda}
\simeq
\frac{y \,{\rm Re}[y_{\Psi}]}{12\pi^2 M_\Psi} \; .
\label{eq:matching}
\end{equation}
Taking $y = -1$, $y_{\Psi} \sim 5$ and $M_\Psi \sim {\cal O}(1)$ TeV
naturally yields the kinetic-mixing parameter 
\(\epsilon \sim 10^{-7}\)
for $m_{W'} \sim 0.01$ GeV, 
consistent with the value required for successful freeze-in production.
The small kinetic-mixing is therefore technically natural,
arising from loop and heavy-mass suppression without fine-tuning. However, we note that in regions of parameter space where $v_\Sigma \gg \Lambda/c_5$, a certain degree of fine-tuning is required in the UV completion to prevent large Yukawa-induced corrections from driving the physical fermion masses away from the bare mass scale $M_\Psi$.

The heavy Dirac lepton carries electric charge \(Q = e\,y\) 
and can be pair-produced at colliders through Drell-Yan processes.
Current LHC searches for such heavy leptons with unit electric charge
exclude masses below about \(1\)~TeV for promptly decaying states~\cite{ATLAS:2023sbu, CMS:2024bni}.

We note that the heavy lepton also carries \(SU(2)_D\) charge
and efficiently annihilates into dark gauge bosons before decaying,
so it does not affect cosmological evolution.
No light remnants remain, preventing contributions to
\(\Delta N_{\rm eff}\) or other precision observables.

This simple UV completion demonstrates that the dimension-5 operator and the resulting millicharge parameter \(\epsilon\)
arise naturally from a renormalizable, gauge-invariant theory,
with loop suppression and heavy-mass scaling that justify
the effective-field-theory treatment used in the main text.

\section{Evolution of the Temperature Ratio \texorpdfstring{$\xi = T_h/T$}{xi = Th/T}}
\label{app:dxidT}

In this and the subsequent three Appendices, we follow the formalism of Refs.~\cite{Aboubrahim:2020lnr,Aboubrahim:2021ycj,Aboubrahim:2021dei,Li:2023nez} closely to review the derivation of the evolution of the temperature ratio \( \xi = T_h /T \) between the hidden and visible sectors with temperature $T_h$ and $T$ respectively.

We start from the continuity equation,
\be
\label{eq:drhodt}
\frac{d\rho}{dt} + 3 H (\rho + p) = 0 \; ,
\ee 
where $\rho = \rho_v + \rho_h$ is the total energy density and $p = p_v + p_h$ is the total pressure. 

Since the hidden and visible sectors feebly interact to each other, one can obtain the energy density evolution equation for each sector by 
\bea
\label{eq:drhovdt}
\frac{d\rho_v}{dt} + 3 H (\rho_v + p_v) &=& -J_h \; , \\ 
\label{eq:drhohdt}
\frac{d\rho_h}{dt} + 3 H (\rho_h + p_h) &=& J_h \; ,
\eea
where the quantity $J_h$ ($-J_h$) represents the source (sink) term of the hidden (visible) sector, encompassing all freeze-in and decay processes that mediate energy exchange between the sectors. 
The explicit expression of $J_h$ is given in Appendix~\ref{app:Jh}. 

Using the chain rule and together with Eqs.~(\ref{eq:drhovdt}) and (\ref{eq:drhohdt}), we obtain 
\be
\label{eq:dThdT}
\frac{dT_h}{dT} = A \, \frac{d\rho_v}{dT} \left(\frac{d\rho_h}{dT_h}\right)^{-1} \; , 
\ee
where 
\be
A \equiv \frac{d\rho_h}{dt} \left(\frac{d\rho_v}{dt}\right)^{-1} = \frac{3 H (\rho_h + p_h) - J_h}{3 H (\rho_v + p_v) + J_h} \; . 
\ee

Now, together with Eq.~(\ref{eq:dThdT}), the evolution equation of the temperature ratio $\xi$ respect to the visible temperature is given by 
\be
\label{eq:dxidT2}
\frac{d\xi}{dT} = \frac{1}{T}\left(-\xi+ \frac{dT_h}{dT} \right)= \frac{1}{T}\left[-\xi + A\, \frac{d\rho_v}{dT} \left(\frac{d\rho_h}{dT_h}\right)^{-1} \right] \; . 
\ee
Hence, knowing $\xi(T)$ as a function of the SM temperature $T$ allows one to compute the hidden temperature as $T_h(T) = T \cdot \xi(T)$.

\section{Pressure, Energy and Entropy Densities}\label{app:rho_p_s}

The visible sector is described by a relativistic plasma of all the SM particles at temperature $T$. 
The corresponding energy density and pressure are given by
\be
\rho_v = \frac{\pi^2}{30} g_{\rm eff}^v  T^4  \; , \qquad
p_v = \frac{1}{3} \rho_v \; , 
\ee
where $g_{\rm eff}^v$ denotes the effective number of relativistic degrees of freedom contributing to the energy density in the visible sector, as obtained from the tabulated results provided by the \texttt{micrOMEGAs} package~\cite{Belanger:2018ccd}.
For the hidden pressure and energy density, we use temperature-dependent integrals given in Ref.~\cite{Hindmarsh:2005ix} to parametrize them which are given by 
\bea
p_h & = & \frac{\pi^2}{90} f_{\rm eff}^h T_h^4  \; , \;\;\;\; f_{\rm eff}^h = f_{
\rm eff}^{W'} + f_{\rm eff}^{Z'} + f_{\rm eff}^{h_D}\; , \\
\rho_h & = & \frac{\pi^2}{30} g_{\rm eff}^h T_h^4  \; , \;\;\;\; g_{\rm eff}^h = g_{
\rm eff}^{W'} + g_{\rm eff}^{Z'} + g_{\rm eff}^{h_D}\; , 
\eea

where 
\bea
\label{eq:feff4i}
f_{\rm eff}^{i} &=& \frac{ 15 g_{i}}{\pi^4} \int_{x_{i}}^{\infty} \frac{ \left(x^2 - x_{i}^2\right)^{3/2} }{e^{x} - 1} dx \, , \\
\label{eq: geff4i}
g_{\rm eff}^i &=& \frac{15 g_{i} }{\pi^4} \int_{x_{i}}^{\infty} \frac{x^2 \left( x^2 - x_{i}^2 \right)^{1/2} }{e^{x} - 1} dx \, , 
\eea
for $i = W', Z', h_D$ and $x_i = m_i/T_h$. 
In the massless limit of $Z'$, the Bose integral is well known
\be
\label{eq:masslessboseintegral}
\int_0^\infty \frac{x^3}{e^x - 1} \, dx = \frac{\pi^4}{15} \; ,
\ee
from which we recover the familiar Stefan-Boltzmann relation for dark radiation:
\be
p_{Z'} = \frac{1}{3} \rho_{Z'} \; 
\; \textrm{ with } \;\; 
\rho_{Z'} = \frac{\pi^2}{30} g_{Z'} T_h^4 \; .
\ee

The total entropy density can be affected by the hidden-sector which is now given by 
\be
\textbf{s} = \frac{2\pi^2}{45} \left(h^v_{\rm eff} T^3 + h^h_{\rm eff} T_h^3 \right), 
\ee
where $h^{v}_{\rm eff}$ and $h^{h}_{\rm eff}$ are entropy degrees of freedoms of the visible and hidden sectors, respectively. 
Here, $h^{v}_{\rm eff}$  is also taken from tabulated results in \texttt{micrOMEGAs} package~\cite{Belanger:2018ccd}, 
while $h^{h}_{\rm eff}$ is calculated using the temperature-dependent integrals given in Ref.~\cite{Hindmarsh:2005ix}, 
\be
h^{h}_{\rm eff} = h^{W'}_{\rm eff} + h^{Z'}_{\rm eff} + h^{h_D}_{\rm eff} \; ,
\ee
where 
\bea
\label{eq:heff4i}
h^{i}_{\rm eff}  &=& \frac{15 g_{i} }{\pi^4} \int_{x_{i}}^{\infty} \frac{ \left( x^2 - x_{i}^2 \right)^{1/2}}{e^{x} - 1} \left( x^2 - \frac{ x_{i}^2 } {4} \right) dx  \; , \;\;\;\; x_i = \frac{m_i}{T_h} \; , \;\;\;\; i = W',Z',h_D \; , 
\eea
with $g_{W'} = 3$, $g_{Z'} = 2$, and $g_{h_D} = 1$. In the case of a massless $Z'$, using Eq.~(\ref{eq:masslessboseintegral}), 
one obtains $h^{Z'}_{\rm eff} = g_{Z'} = 2$.

The Hubble expansion rate is also affected by the presence of the hidden sector which is now given by 
\be
H^2 = \frac{8\pi G_N}{3} \left( \rho_v +  \rho_h \right) \; ,
\ee
where $G_N$ is Newton's gravitational constant.

\section{The Source Term \texorpdfstring{$J_h$}{Jh}}
\label{app:Jh}

The function $J_h$ accounts for the source terms of the hidden sector from all possible processes that exchange energy between the SM and the hidden sectors which is given by
\bea
\label{eq:Jh}
\begin{aligned}
J_h &\simeq \sum_{f}\Big\{n_f^2\left[2 J_{f\bar{f} \to W^pW^m}(T)   + J_{f\bar{f} \to h_D}(T) \right]   \\ 
& \hspace{1cm} +\, 2 n_{Z} J_{Z\to W^pW^m}(T) +  2 n_{h} J_{h\to W^pW^m}(T)   \\
& \hspace{1cm} +\, 2 n_{\gamma^*} J_{\gamma^*\to W^pW^m}(T)  - n_{h_D} J_{h_D \to f\bar{f}} (T_h) \Big\} \; ,
\end{aligned}
\eea
where \(n_i\) denotes the number density of species \(i\), and we have also included the plasmon decay process $\gamma^* \to W^p W^m$.
The $J$-functions appeared on the right-hand side in Eq.~(\ref{eq:Jh}) are defined by 
\bea
n_a n_b J_{a b \to cd} &=& \frac{g_a g_b T}{128 \pi^4}\int_{s_0}^{\infty} ds \, \sigma_{cd \to ab}(s) s (s - s_0) K_2(\sqrt{s}/T) \; , \\
n_f^2 J_{f \bar{f} \to h_D} &=& \frac{T}{32 \pi^4}\int_{s_0}^{\infty} ds \, \sigma_{f \bar{f} \to h_D}(s) s (s - s_0) K_2(\sqrt{s}/T) \; , \\
 n_{a} J_{a \to bc} &=& n_{a} m_{a} \Gamma_{a \to b c }\;\; , \\
 n_{\gamma^*} J_{\gamma^* \to W^p W^m } &=& 
 n_{\gamma_\ell^*} m_l \langle\Gamma\rangle_{\gamma_\ell^{*}\to W^p W^m} +
 n_{\gamma_t^*} m_t \langle\Gamma\rangle_{\gamma_t^{*}\to W^p W^m} 
 \; , 
\label{eq:Jh-plasmon}
\eea
where $s_0$ is the minimum square of the center-of-mass energy.
For $ab\to cd$, $s_0 = {\rm max}((m_a + m_b)^2, (m_c + m_d)^2)$; while for $f \bar f \to h_D$, $s_0 = 4 m_f^2$.
For the longitudinal and transverse plasmon decay rates in the right-hand side of Eq.~(\ref{eq:Jh-plasmon}), see Appendix~\ref{app:plasmondecay}.

\section{Coupled Boltzmann Equations}
\label{app:CoupledBoltzmann}

The number density of particles \(n_i\) in the hidden sector can be derived from the following coupled Boltzmann equations:
\bea
\label{eq:BZeqi}
\frac{dn_{i}}{dt} + 3 H n_{i} &=& {\mathcal{R}}_{i}(T,T_h) \; , \;\;\;\; i = W', Z', h_D \; , 
\eea
where $t$ is the evolution time and we assume $n_{W'} = n_{W^m} = n_{W^p}$ for symmetric dark matter. In the right hand side of Eq.~(\ref{eq:BZeqi}), 
${\mathcal{R}}_{i}(T,T_h)$, 
represents the number of interactions per unit volume per unit time that produce $i$ with  \( i= W', Z' \) and $h_D$, respectively. Explicitly they are given by
\bea
\label{eq:CWp}
 {\mathcal{R}}_{W'} (T,T_h) &=& \sum_{f\in\rm{SM}} \biggl[ n_f^2 \langle \sigma v \rangle_{f\bar{f} \to W^p W^m}(T) \biggr] + n_{Z} \langle \Gamma \rangle_{Z \to W^pW^m} (T) + n_{h} \langle \Gamma \rangle_{h \to W^pW^m} (T)  \nonumber \\
 &&  \hspace{1cm}    + \, n_{\gamma^*} \langle \Gamma \rangle_{\gamma^* \to W^pW^m} (T)  + n_{h_D}  \langle \Gamma \rangle_{h_D \to W^p W^m} (T_h) \nonumber \\
 &&  \hspace{1cm}  
 + \,  n_{h_D}^2 \ \langle \sigma v \rangle_{h_D h_D \to W^p W^m}(T_h) 
 - \, n_{W'}^2 \, \biggl[ \langle \sigma v \rangle_{W^p W^m \to h_D}(T_h) \bigr.
  \nonumber \\
 &&  \hspace{1cm}
     \bigl. + \, \langle \sigma v \rangle_{W^p W^m \to Z' Z'}(T_h) +      \langle \sigma v \rangle_{W^p W^m \to h_D h_D}(T_h) \biggr] 
 \; , \\
 \label{eq:CZp}
{\mathcal{R}}_{Z'}(T,T_h) &=& n_{W'}^2  \langle \sigma v \rangle_{W^pW^m \to Z' Z'}(T_h)   
\; ,
\\
\label{eq:Ch_D}
{\mathcal{R}}_{h_D}(T,T_h) &=&  \sum_{f\in\rm{SM}} \bigg[ n_f^2 \langle \sigma v \rangle_{f\bar{f} \to h_D}(T) \biggr] + \, n_{W'}^2  \biggl[ \langle \sigma v \rangle_{W^pW^m \to h_D h_D}(T_h)  \nonumber \\
&& \hspace{1cm}  + \,  \langle \sigma v \rangle_{W^p W^m \to h_D}(T_h) \biggr] -   n_{h_D}^2  \langle \sigma v \rangle_{h_D h_D \to W^p W^m}(T_h)  \nonumber \\ 
&& \hspace{1cm}  - \, n_{h_D}  \biggl[ \langle \Gamma \rangle_{h_D \to W^p W^m} (T_h) 
+ \sum_{f\in\rm{SM}} \langle \Gamma \rangle_{h_D \to f\bar{f}} (T_h)
\bigg] \; .
\eea
Here, $\langle \sigma v \rangle$ and $\langle \Gamma \rangle$ denote the thermally averaged cross section and decay width, respectively.
For computational simplicity, one can use the principle of detailed balance to obtain the following relation for the thermally averaged cross sections:
\bea
n_f^2 \langle \sigma v \rangle_{f\bar{f} \to W^p W^m}(T) &\equiv& (n_{W'}^{eq})^2 \, \langle \sigma v \rangle_{ W^p W^m \to f\bar{f}  }(T) \; ,
\eea
where $n_{W'}^{eq}$ is the corresponding number density of $W'$ in equilibrium, which is given by 
\be
\label{eq:nWprimeEquilibrium}
n_{W'}^{eq} = \frac{g_{W'}}{2\pi^2} m_{W'}^2 T K_2(m_{W'}/T) \;.
\ee
Here (and below) 
$K_i$ is the modified Bessel function of the second kind of degree $i$.

The thermally averaged cross section for a $ab \to cd$ process in Eqs.~(\ref{eq:CWp}), (\ref{eq:CZp}) and  (\ref{eq:Ch_D}) is given by~\cite{Chu:2011be}:
\bea
\label{eq:avxsec22}
\begin{aligned}
\langle \sigma v \rangle_{a b \to c d}(T) &= \frac{1}{8 \, {c_{ab}} \, m_a^2 m_b^2 \,T K_2(m_a/T) K_2(m_b/T)}  \\
& \times \int_{(m_a + m_b)^2}^{\infty} ds \, \sigma_{a b\to c d}(s)  \frac{1}{\sqrt{s}} \left[ \left(s - m_a^2 - m_b^2\right)^2 - 4 m_a^2 m_b^2 \right] K_1(\sqrt{s}/T) \; ,
\end{aligned}
\eea
where $c_{ab}$ is a combinatorial factor, taking the value 2 (1) when $a$ and $b$ are identical (nonidentical), and $s$ denotes the square of the center-of-mass energy.
The cross sections $\sigma(s)$ for various processes are provided in Appendix~\ref{app:xsec_decay}.

The thermally averaged cross section for an $a\bar{a} \to b$ process is given by:
\be
\label{eq:avxsec21}
\langle \sigma v \rangle_{a\bar{a} \to b}(T) = \frac{1}{8 m_a^4  \,T K^2_2(m_a/T) }\int_{4 m_a^2}^{\infty} ds \, \sigma_{a \bar{a}\to b}(s) \sqrt{s} \left[ s - 4 m_a^2 \right] K_1(\sqrt{s}/T) \; .
\ee

The thermally averaged decay width of $Z$, $h$ and $h_D$ bosons 
in Eqs.~(\ref{eq:CWp}), (\ref{eq:CZp}) and (\ref{eq:Ch_D}) 
is given by: 
\be
\langle \Gamma \rangle_{a \to b c} (T) = \Gamma_{a \to b c} \frac{K_1(m_a/T)}{K_2(m_a/T)} \; . 
\ee

The thermally averaged rate of plasmon decay ($\gamma^* \to W^p W^m$) in Eq.~(\ref{eq:CWp}) is provided in Appendix~\ref{app:plasmondecay}.

For the computation of the relic density, it is more convenient to work directly with the comoving number densities (yields) of DM and mediators, defined as $Y_{W'} = n_{W'}/{\textbf s}$, $Y_{Z'} = n_{Z'}/{\textbf s}$ and $Y_{h_D} = n_{h_D}/{\textbf s}$, where ${\textbf s}$ is the total entropy density given in 
Appendix~\ref{app:rho_p_s}. 

By changing variables to the yield, the coupled Boltzmann equation in Eqs.~(\ref{eq:BZeqi}), becomes
\bea
\label{eq:dYidT0}
 -\frac{\textbf{s}\, dY_{i}}{d T} \left[ \frac{  3H(\rho_v + p_v) + J_h]}{d\rho_v/dT} \right] + Y_{i}\left(\frac{d\textbf{s}}{dt} + 3 H \textbf{s}\right)&=&  {\mathcal{R}}_{i}(T,T_h) \; , \;\;\;\; i = W', Z', h_D \; ,
\eea
where we used the energy density evolution equation in (\ref{eq:drhovdt}) to change time to temperature, namely, 
\be
\frac{dT}{dt} = -\left[3H(\rho_v + p_v) + J_h \right] \times \left( \frac{d\rho_v}{dT} 
\right)^{-1} \; .
\ee
Assuming the total entropy, $\textbf{S} =  a^3\textbf{s}$ where $a$ is the scale factor of the universe, is conserved which gives the continuity equation for entropy density 
\be 
\frac{d \textbf{s}}{dt} + 3 H \textbf{s} = 0 \; ,
\ee
we can rewrite Eq.~(\ref{eq:dYidT0})
as 
\bea
\label{eq:dYidT2}
\frac{dY_{i}}{d T} &=& -\frac{1}{\textbf s} \left[\frac{d\rho_v/dT}{3H(\rho_v + p_v) + J_h} \right] \times  {\mathcal{R}}_{i}(T,T_h) \; , \;\;\;\; i = W', Z', h_D  \; . 
\eea

The system of differential equations in Eqs.~(\ref{eq:dxidT2}) and (\ref{eq:dYidT2})
can be solved numerically to track the evolution of the yields of the three dark species. In particular, integrating the DM yield to the present-day temperature yields $Y_{W'}^0$.

The DM relic density is then determined by
\be
\Omega_{\mathrm{DM}} h^2 = \frac{m_{W'} Y_{W'}^0 {\textbf s}_0 h^2}{\rho_c} \; ,
\ee 
where $\rho_c$ is the critical energy density required to close the universe, $ {\textbf s}_0 $ is the current entropy density, and \( h \) is the dimensionless Hubble parameter defined as
\begin{equation}
h \equiv \frac{H_0}{100~\mathrm{km\,s^{-1}\,Mpc^{-1}}} \; ,
\end{equation}
with \( H_0 \) being the present-day Hubble expansion rate.
According to the latest results from the Planck Collaboration, the DM relic density is measured to be 
~\cite{Planck:2018vyg}
\begin{equation}
\Omega_{\mathrm{DM}} h^2 = 0.120 \pm 0.001 \; .
\end{equation}
We will use $h = 0.678$ in our analysis.

\section{Relevant Cross Sections and Decay Widths}
\label{app:xsec_decay}

In this Appendix, we provide the relevant cross sections and decay widths to the freeze-in mechanism.
\begin{figure}[htbp!]
        \centering
	\includegraphics[width=0.65\textwidth]{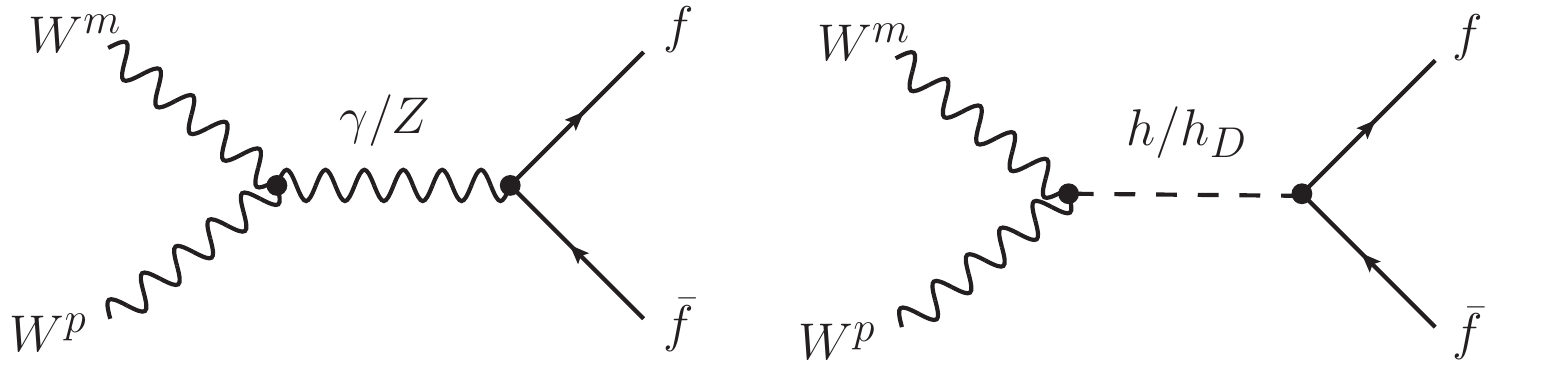}
	\caption{ \label{fig:feyndiag_WmWpff} Feynman diagrams for the process of dark matter annihilation into a pair of the SM fermions.}
\end{figure}

\subsection{\texorpdfstring{$W^p W^m \to \bar{f}f$}{Wp Wm to f-bar f}}\label{app:xsec_decay_1}

The annihilation of dark matter into a pair of SM fermions via an s-channel process mediated by neutral gauge boson (a photon and $Z$ boson) and scalar bosons (Higgs and dark Higgs) is illustrated in Fig.~\ref{fig:feyndiag_WmWpff}. 
The interference between the scalar-boson-mediated and gauge-boson-mediated diagrams is proportional to \( (u - t) \), which vanishes upon integration over the full phase space. This cancellation arises from the orthogonality of scalar and vector current structures, which project onto different partial waves specifically, s-wave for scalars and p-wave for vectors. Consequently, the interference terms are odd under parity and integrate to zero over symmetric angular distributions, even in the presence of nonzero fermion masses.

The cross section from gauge-boson-mediated diagrams is given as 
\bea
\sigma(W^p W^m \to \bar{f}f)_{(\gamma,Z)} &=&  \frac{N_f g_D^2 \beta_{W'} \beta_f }{108 \pi \, m_{W^\prime}^2 } \nonumber\\
&& \hspace{-3.5cm} \times \Bigg\{ e^2 Q_f^2  \kappa_{11}  \left(1 + \frac{2 m_f^2}{s} \right) 
+ \frac{2 s (s - m_Z^2) \kappa_{12}}{\left( s - m_Z^2 \right)^2 + m_Z^2 \Gamma_Z^2} \left[ e Q_f v_Z^f  \left( 1 + \frac{2 m_f^2}{s} \right) \right]  \\
&& \hspace{-3.5cm} + \frac{s^2  \kappa_{22} }{\left( s - m_Z^2 \right)^2 + m_Z^2 \Gamma_Z^2} \left[\vert v_Z^f \vert ^2 \left( 1 + \frac{2 m_f^2}{s} \right) + |a_Z^f|^2 \left( 1 - \frac{4 m_f^2}{s} \right) \right] 
\Bigg\} \, ,  \nonumber 
\eea
where $\beta_{W'} = \sqrt{ 1- \frac{4 m_{W^\prime}^2}{s} }$ and $\beta_{f} = \sqrt{ 1- \frac{4 m_{f}^2}{s} }$ are velocities of the dark matter and fermions, respectively, $s$ is the Mandelstam variable representing the square of the total energy in the center-of-mass frame, 
$m_{W^\prime}=m_{W^m}=m_{W^p}$, $N_f = 1 \; \mathrm{or} \; 3$ is the number of color for the fermion $f$, and 
\be
\label{eq:kappaij}
\kappa_{ij}  =    \kappa_{i} \kappa_{j} \left(1 + \frac{3 m_{W^\prime}^2}{s} \right) \, , 
\ee
where $\kappa_1$ and $\kappa_2$ are given by Eq.~(\ref{eq:kappa123}).

The cross section from scalar-boson-mediated diagrams is given as
\bea
\begin{aligned}
\sigma(W^p W^m \to \bar{f}f)_{(h,h_D)} &= \frac{N_f  g_D^2 (m_f^2/v^2) \cos^2\beta \sin^2\beta }{32 \pi \, m_{W^\prime}^2 } \frac{\beta_f^3 }{\beta_{W'} }   \\
&\times \frac{s^2 \left(m_h^2 - m_{h_D}^2 \right)^2}{\left(s-m_h^2\right)^2 \left(s- m_{h_D}^2 \right)^2} \left(1- \frac{4 m_{W'}^2}{s} +  \frac{12 m_{W'}^4}{s^2}\right) \, ,
\end{aligned}
\eea
where $v = g \,m_W/2$.

The total cross section is then given by 
\be
\sigma(W^p W^m \to \bar{f}f) = \sigma(W^p W^m \to \bar{f}f)_{(h,h_D)} +\sigma(W^p W^m \to \bar{f}f)_{(\gamma,Z)} \; .
\ee

\subsection{\texorpdfstring{$Z \to W^p W^m $}{Z --> Wp Wm}}
\label{app:xsec_decay_2}

The decay rate of the SM $Z \to W^p W^m$ process is given as 
\bea
\begin{aligned}
\Gamma(Z \to W^p W^m) & = \frac{ g_D^2 \kappa_{2}^2 m_{Z}}{48 \pi} \left(1 - \frac{4 m_{W^\prime}^2}{m_{Z}^2}\right)^{3/2} \left(3 + \frac{m_{Z}^2} {m_{W^\prime}^2} \right) . 
\end{aligned}    
\eea
where the coupling ${\kappa}_{2}$ is given in Eq.~(\ref{eq:kappa123}),

\begin{figure}[htbp!]
        \centering
	\includegraphics[width=0.75\textwidth]{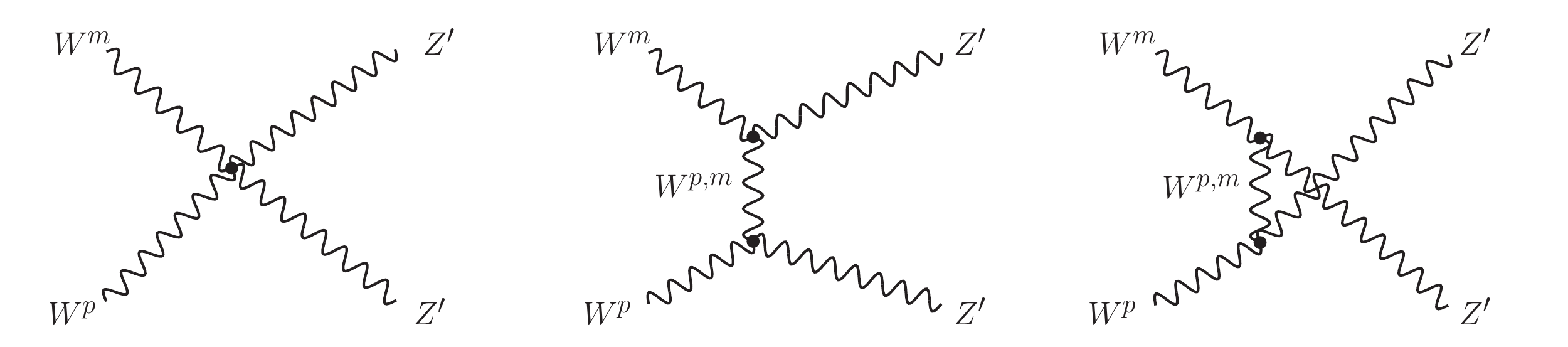}
	\caption{ \label{fig:feyndiag_WmWpZpZp} Feynman diagrams for the process of dark matter annihilation into di-$Z'$ bosons.}
\end{figure}

\subsection{\texorpdfstring{$W^p W^m \to Z^\prime Z^\prime$}{Wp Wm --> Z' Z'}}\label{app:xsec_decay_3}

The annihilation of dark matter into massless di-$Z'$ bosons is depicted in Fig.~\ref{fig:feyndiag_WmWpZpZp}, which mimics the SM process \( W^+ W^- \to \gamma \gamma \).
The total cross section for this process is given as
\bea
\label{eq:WpWmtoZpZp}
\begin{aligned}
\sigma(W^p W^m \to Z' Z') & = \frac{2 g_D^4}{3 \pi m_{W^\prime}^2 \beta_{W'}^2 } \Bigg[ \beta_{W'} \left(\frac{1}{3} + \frac{m_{W^\prime}^2}{4 s}+ \frac{ m_{W^\prime}^4}{s^2} \right)  \\
& + \;\;   \frac{ m_{W^\prime}^4}{s^2} \left( 1 -  \frac{2 m_{W^\prime}^2}{s} \right) \log \left(\frac{1- \beta_{W'}}{1+\beta_{W'} }\right) 
\Bigg] \, . 
\end{aligned}
\eea 

For a massless dark photon \( Z' \), the sum over its physical transverse polarizations introduces a dependence on an auxiliary lightlike vector \( n^\mu \):
\begin{equation}
\sum_{\lambda=1,2} \epsilon_\mu^{(\lambda)}(k) \epsilon_\nu^{*(\lambda)}(k)
= -g_{\mu\nu} + \frac{k_\mu n_\nu + k_\nu n_\mu}{k \cdot n}
\end{equation}
where \( k^\mu \) is the dark photon momentum and \( n^\mu \) satisfies \( n^2 = 0, k \cdot n \neq 0 \). This construction projects out the unphysical longitudinal mode, and despite the intermediate \(n^\mu\)-dependence, the final cross section in Eq.~(\ref{eq:WpWmtoZpZp}) is gauge invariant and independent of this choice.

Similar expressions can be derived for the processes $W^p W^m \to \gamma \gamma$ and $W^p W^m \to Z^\prime \gamma$. However, these contributions are suppressed by the kinetic-mixing parameter \(\epsilon\) relative to the dominant $W^p W^m \to Z^\prime Z^\prime$ channel.

\begin{figure}[htbp!]
        \centering
	\includegraphics[width=0.6\textwidth]{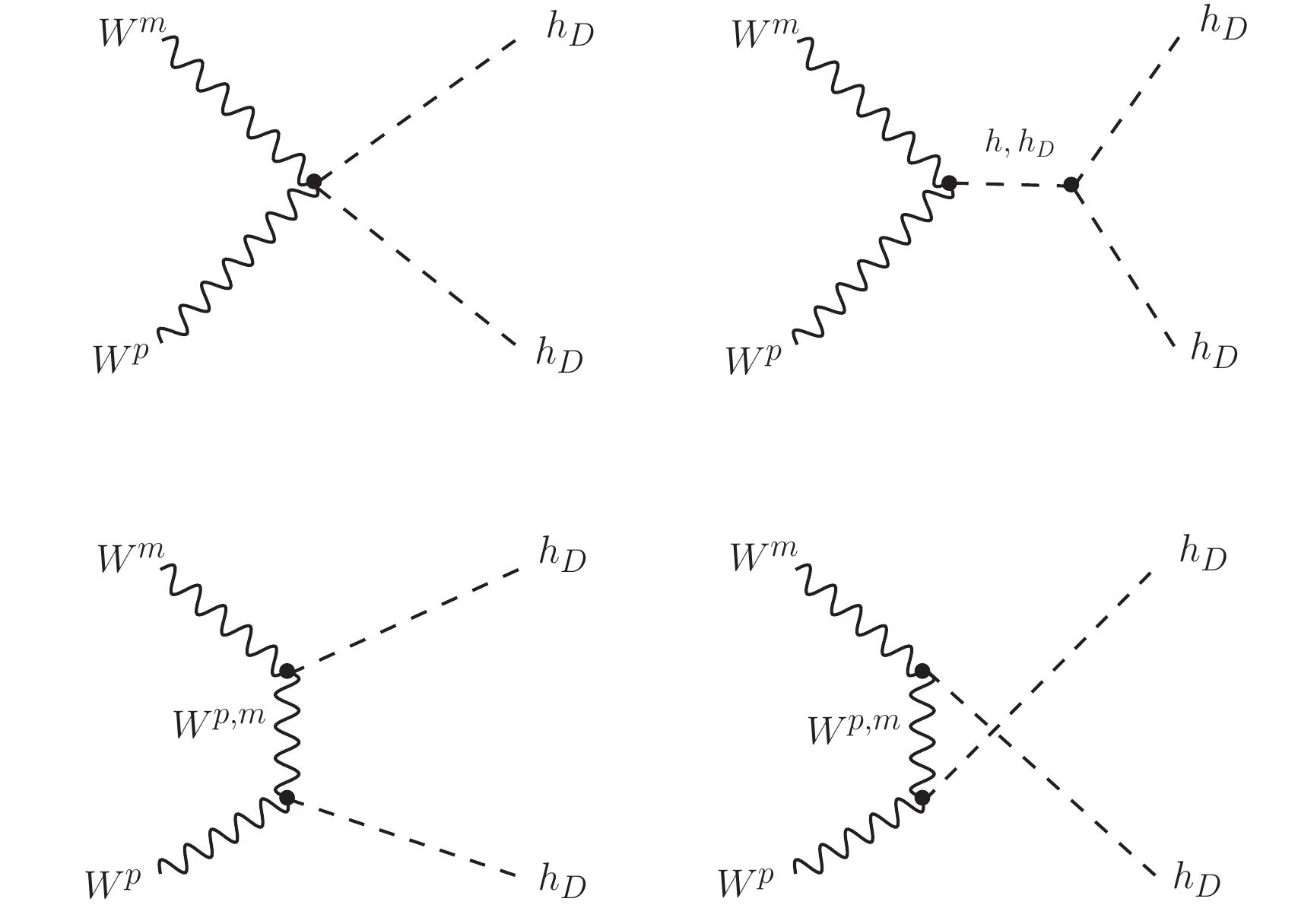}
	\caption{ \label{fig:feyndiag_WmWphDhD} Feynman diagrams for the process of dark matter annihilation into a pair of dark scalar boson $h_D$.}
\end{figure}

\subsection{\texorpdfstring{$W^p W^m \to h_D h_D$}{Wp Wm --> hD hD}}\label{app:xsec_decay_4}

We consider the process 
\be
W^p(p_1,\lambda_1) W^m(p_2, \lambda_2) \to h_D(k_1) h_D (k_2) \; ,
\ee
where $\lambda_{1,2} = 0, \pm 1$ denotes the helicity of the incoming $W^{p,m}$. The Feynman diagrams for this process 
in unitary gauge is shown in Fig.~\ref{fig:feyndiag_WmWphDhD}. 

The helicity amplitude of the process is given by 
\be
\label{eq:helicityamplitudes}
{\cal M}^{\lambda_1\lambda_2}_{W^p W^m \to h_D h_D} = {\cal M}_{\mathrm{4pt}}^{\lambda_1\lambda_2}  + {\cal M}_{h}^{\lambda_1\lambda_2} 
+ {\cal M}_{h_D}^{\lambda_1\lambda_2} + {\cal M}_{W',t}^{\lambda_1\lambda_2} + {\cal M}_{W',u}^{\lambda_1\lambda_2} \;  ,
\ee
where 
\bea
\begin{aligned}
{\cal M}_{\mathrm{4pt}}^{\lambda_1\lambda_2} &= 2 g_D^2  \cos^2\beta  \, \epsilon_1 \cdot \epsilon_2 \; , \\
{\cal M}_{h}^{\lambda_1\lambda_2} &= -2 g_D m_{W'}  \lambda_{hh_Dh_D} \sin\beta \, \frac{ \epsilon_1 \cdot \epsilon_2 }{(k_1 + k_2)^2 - m_h^2} \; ,\\
{\cal M}_{h_D}^{\lambda_1\lambda_2} &= 2 g_D m_{W'} \lambda_{h_Dh_Dh_D} \cos\beta\,\frac{\epsilon_1\cdot \epsilon_2 }{(k_1 + k_2)^2 - m_{h_D}^2} \; ,\\
{\cal M}_{W',t}^{\lambda_1\lambda_2} &= 4 g_D^2 \cos^2\beta \frac{m_{W'}^2 \epsilon_1 \cdot \epsilon_2 - (k_2 - p_2)\cdot \epsilon_1 \, (k_2 - p_2)\cdot \epsilon_2 }{(k_2 - p_2)^2 - m_{W'}^2} \; , \\
{\cal M}_{W',u}^{\lambda_1\lambda_2} &= 4 g_D^2 \cos^2\beta \frac{m_{W'}^2 \epsilon_1 \cdot \epsilon_2 - (k_1 - p_2)\cdot \epsilon_1 \,(k_1 - p_2) \cdot \epsilon_2 }{(k_1 - p_2)^2 - m_{W'}^2} \; , 
\end{aligned}
\eea
with $\epsilon_i  =  \epsilon_i(p_i,\lambda_i), \;i = 1,2$ are the two polarization vectors of $W^p$ and $W^m$, and
\bea
\begin{aligned}
\lambda_{hh_Dh_D} &=  6 \sin\beta \cos\beta \left( \lambda_\phi v \sin\beta  - \lambda_{\Sigma} v_\Sigma \cos\beta \right) \\ 
 &  + 2 \, \lambda_{\Phi \Sigma} \left[ v \cos\beta \left(\frac{1}{2}\cos^2\beta - \sin^2\beta \right) 
 - v_\Sigma \sin\beta \left(\frac{1}{2}\sin^2\beta - \cos^2\beta\right) \right] \, , \\ 
 \lambda_{h_Dh_Dh_D} & = 3 \Big[ 2 \lambda_\phi v \sin^3\beta + 2 \lambda_{\Sigma} v_\Sigma \cos^3 \beta 
 + \lambda_{\Phi \Sigma} \sin\beta \cos\beta \left( v \cos \beta + v_\Sigma \sin\beta\right) \Big] \, . 
 \end{aligned}
\eea

In the center-of-mass system, the momenta are given as 
\bea
\begin{aligned}
p_1^\mu &= \frac{\sqrt{s}}{2}(1, 0, 0, \beta_{W'}) \; , \\ 
p_2^\mu &= \frac{\sqrt{s}}{2}(1, 0, 0, -\beta_{W'}) \; ,\\
k_1^\mu &= \frac{\sqrt{s}}{2}(1, \beta_{h_D} \sin \theta_{\rm cm}, 0, \beta_{h_D} \cos \theta_{\rm cm}) \; ,\\
k_2^\mu &= \frac{\sqrt{s}}{2}(1, -\beta_{h_D} \sin \theta_{\rm cm}, 0, -\beta_{h_D} \cos \theta_{\rm cm}) \; ,
\end{aligned}
\eea
where $\beta_{h_D} = \sqrt{1 - \frac{4 m_{h_D}^2}{s} }$ is the velocity of $h_D$ and $\theta_{\rm cm}$ is the angle between the momenta $\vec{k}_1$ and $\vec{p}_1$ in the center-of-mass frame.

The explicit polarization vectors of $W^{p,m}$ can be written down as 
\bea
\begin{aligned}
\epsilon^\mu_1(p_1,\lambda_1 = \pm 1) &= \frac{1}{\sqrt{2}} (0, 1, \pm i, 0) \; ,  \\
\epsilon^\mu_2(p_2,\lambda_2 = \pm 1) &= \frac{-1}{\sqrt{2}} (0, 1, \mp i, 0) \; ,  \\
\epsilon^\mu_1(p_1,\lambda_1 = 0) &= \frac{1}{m_{W'}} \frac{\sqrt{s}}{2}(\beta_{W'}, 0, 0, 1) \; , \\
\epsilon^\mu_2(p_2,\lambda_2 = 0) &= \frac{1}{m_{W'}} \frac{\sqrt{s}}{2}(\beta_{W'}, 0, 0, -1) \; .
\end{aligned}
\eea

Altogether, nine helicity configurations are possible. With the exception of $ \lambda_1 = \lambda_2 = 0 $, the other eight helicity amplitudes satisfy the symmetry relations given below:
\bea
\begin{aligned}
{\cal M}^{++}_{W^p W^m \to h_D h_D} &= {\cal M}^{--}_{W^p W^m \to h_D h_D} \; , \\
{\cal M}^{-+}_{W^p W^m \to h_D h_D} &= {\cal M}^{+-}_{W^p W^m \to h_D h_D} \; , \\
{\cal M}^{+0}_{W^p W^m \to h_D h_D} &= {\cal M}^{0+}_{W^p W^m \to h_D h_D} = {\cal M}^{-0}_{W^p W^m \to h_D h_D} = {\cal M}^{0-}_{W^p W^m \to h_D h_D} \; .
\end{aligned}
\label{eq:helicityamplitudesSymmetry}
\eea
The differential cross section can be obtained by 
\be 
\label{eq:xsecWpWmhDhD}
\frac{d\sigma ({W^p W^m \to h_D h_D})}{d\cos{\theta_{\rm cm}}} = \frac{1}{32 \pi s} \frac{\beta_{h_D}}{\beta_{W'}} \frac{1}{2 \times 9} \sum_{\lambda_1,\lambda_2} \Big| {\cal M}^{\lambda_1\lambda_2 }_{W^p W^m \to h_D h_D} \Big|^2 \, ,
\ee
where the factors of $2$ and $9$ in the denominator of Eq.~(\ref{eq:xsecWpWmhDhD}) account for identical particles of dark scalars in the final state and average of $W^{p,m}$ polarization in the initial state, respectively.  
The explicit forms of the nine helicity amplitudes in Eq.~(\ref{eq:helicityamplitudes}) and the resulting cross section are cumbersome and provide little additional insight.
They will be omitted here.

\subsection{\texorpdfstring{$ h_D h_D \to W^p W^m $}{hD hD --> Wp Wm}}\label{app:xsec_decay_5}

The cross section of $h_D h_D \to W^p W^m $ can be obtained by its relation to $W^p W^m \to h_D h_D$ process, which is given as
\be 
\label{eq:xsechDhDWpWm}
\sigma({ h_D h_D \to W^p W^m }) = 18 \left( \frac{\beta_{W'}}{\beta_{h_D}} \right)^2 \times \sigma({W^p W^m \to h_D h_D}) \; .
\ee

\subsection{\texorpdfstring{$W^p W^m \to h_D$}{W W --> hD}}
\label{app:xsec_decay_6}

The cross section of $W^p W^m \to h_D$ can be given as 
\be
\sigma_{W^p W^m \to h_D} =\frac{\pi  g_D^2 m_{h_D}^2 \cos^2\beta}{9 \beta_{W'} s } \left(  - \, 4 + \frac{12 m_{W'}^2}{m_{h_D}^2} + \frac{m_{h_D}^2}{m_{W'}^2} \right) \delta(s - m_{h_D}^2) \; . 
\ee

\subsection{\texorpdfstring{$h_D \to W^p W^m$}{hD --> Wp Wm}}
\label{app:xsec_decay_7}

If $m_{h_D} > 2 m_{W'}$, the dark scalar $h_D$ can decay into $W^p W^m$. 
The decay width of this process is given as 
\be
\Gamma(h_D \to W^p W^m) = \frac{ g_D^2 m_{h_D} \cos^2\beta }{16 \pi} \sqrt{1 - \frac{4 m_{W'}^2}{m_{h_D}^2}} \left(  - 4 + \frac{12 m_{W'}^2}{m_{h_D}^2} + \frac{m_{h_D}^2}{m_{W'}^2} \right) \; . 
\ee

\subsection{\texorpdfstring{$\bar{f} f \to h_D$}{fbar f --> hD}}\label{app:xsec_decay_8}

The cross section of $\bar{f} f \to h_D$ process is given as
\be
\sigma(\bar{f} f \to h_D) = \frac{\pi \, m_f^2 \, m_{h_D}^2 \sin^2\beta } {2 N_f v^2 \beta_f s} \left( 1 - \frac{4 m_f^2}{m_{h_D}^2}\right) \delta(s - m_{h_D}^2) \; . 
\ee

\subsection{\texorpdfstring{$h \to \bar{f} f$ and $h_D \to \bar{f} f$}{h --> fbar f and hD --> fbar f}}\label{app:xsec_decay_9}

If kinematically allowed, scalar bosons can decay into a pair of SM fermions. The decay width of these processes can be given by 
\bea
\Gamma(h \to \bar{f} f) &=& \frac{N_f  \, m_{f}^2 \, m_{h} \cos^2\beta }{16 \pi v^2} \left(1 - \frac{4 m_{f}^2}{m_{h}^2} \right)^{3/2} \; , \\
\Gamma(h_D \to \bar{f} f) &=& \frac{N_f  \, m_{f}^2\, m_{h_D} \sin^2\beta }{16 \pi v^2} \left(1 - \frac{4 m_{f}^2}{m_{h_D}^2} \right)^{3/2} \; . 
\eea

\section{Plasmons Decay into Vector Dark Matter}\label{app:plasmondecay}

The decay of photons and plasmons into neutrino pairs was first calculated in~\cite{Braaten:1993jw} to account for the energy loss in a stellar plasma due to the plasma process. Recently, this calculation was extended to the fermionic millicharged DMs in \cite{Dvorkin:2019zdi}. Here we extend further to the plasmon decay into vectorial millicharged DMs. 
The effective matrix element for the process $\gamma^{*}(k) \to W^m (p_1) W^p (p_2)$ is given by 
\bea
\label{eq:Mplasmondecay}
\begin{aligned}
i {\cal M}_{\gamma^{*} \to W^m W^p} &= i  g_{D}  \kappa_1 \tilde{\epsilon}_\mu(k) \epsilon^{*}_{\rho}(p_1) \epsilon^{*}_{\sigma}(p_2)
 \Big( (p_1 - p_2)^\mu g^{\rho \sigma} - p_1^{\sigma} g^{\mu \rho} + p_2^{\rho} g^{\mu\sigma} \Big)  \; , 
\end{aligned}
\eea
where the millicharge couplings $ \kappa_1$ are given in Eq.~(\ref{eq:kappa123}), $K^\mu = (\omega(k), \Vec{k})^\mu$ is the 4-momentum of photon, $p_1^\mu = (E_1, \Vec{p}_1)^\mu$ is the 4-momentum of $W^m$ and $p_2^\mu = (E_2, \Vec{p}_2)^\mu$ is the 4-momentum of $W^p$.
In Eq.~(\ref{eq:Mplasmondecay}), $\epsilon^{*}_\rho (p_1)$ ($\epsilon^{*}_\sigma (p_2) $) is the polarization vector for $W^m$ ($W^p$) and $\tilde{\epsilon}_\mu(k)$ is the dressed polarization vector for the plasmon which are given by \cite{Braaten:1993jw}, 
\bea
\tilde{\epsilon}_L^\mu(k) &=& \frac{\omega_\ell(k)}{k} \sqrt{Z_\ell(k)} (1, \Vec{0})^\mu  
\;\;\;\;\; \text{(longitudinal mode) \, , } \\
\tilde{\epsilon}_\pm^\mu(k) &=& \sqrt{Z_t(k)} (0, \Vec{\varepsilon}_\pm)^\mu  
\;\;\;\;\;\;\;\;\;\;\;\; \text{(transverse mode) \, , } 
\eea
with $\omega_\ell(k)$ [$\omega_t(k)$] and $Z_\ell(k)$ [$Z_t(k)$]  are the dispersion relation and the wave function renormalization factor for the longitudinal (transverse) mode, respectively\footnote{See Ref. \cite{Braaten:1993jw} for the expressions of $\omega_{\ell,t}(k)$ and $Z_{\ell,t}(k)$. }. 

Squaring the matrix element in Eq.~(\ref{eq:Mplasmondecay}) and then summing over the polarizations, one finds that only the diagonal terms ($LL$, $++$, and $--$) contribute, as one would expect.
The squared matrix element for the  $LL$ configuration is given by 
\bea
\begin{aligned}
\vert{\cal M}\vert^2_{LL} & = \frac{g_{D}^2 \kappa_1^2 Z_{\ell} \omega_\ell^2}{k^2}   \Big[ 4 K_\ell\cdot p_1 + 12 E_1 (E_1 - \omega_\ell) + \omega_\ell^2 \\ & \hspace{2.0cm} 
-\, \frac{2K_\ell\cdot p_1}{m_{W^\prime}^2} \Big( K_\ell\cdot p_1 + 2 E_1(E_1 - \omega_\ell) \Big) \Big] \; ,
\end{aligned}
\eea
where $m_\ell^2 = ( \omega_\ell(k)^2 - k^2 )$ is the effective plasmon mass for the longitudinal mode and  $K_\ell\cdot p_1 = E_1\omega_\ell - k p_1\cos\theta$, with $\theta$ is the angle between the plasmon and the DM momenta.  

In contrast, the squared matrix element for the \(++\) and \(--\) helicity configurations is given by
\bea
\sum_{TT = ++ , --} \vert{\cal M}\vert^2_{TT} =  4 g_{D}^2 \kappa_1^2  Z_{t}  
 \Big[-2 K_t\cdot p_1 + 3 p_1^2 \sin^2\theta + \frac{K_t\cdot p_1}{m_{W^\prime}^2}  \left(K_t\cdot p_1 - p_1^2 \sin^2\theta \right)\Big] 
 \, , \;
\eea
where $m_t^2 = (\omega_t (k)^2 - k^2)$ is the effective plasmon mass for the transverse mode and $K_t\cdot p_1 = E_1\omega_t - k p_1\cos\theta$.

The thermally averaged decay rate can be obtained by 
\bea
\begin{aligned}
n_{\gamma^*} \langle\Gamma\rangle_{\gamma^{*}\to W^p W^m}  &= \int \frac{d^3k}{(2\pi)^3 2\omega} \frac{d^3p_1}{(2\pi)^3 2E_1} \frac{d^3p_2}{(2\pi)^3 2E_2} f_B (\omega(k)) \\
&\hspace{0.6cm} \times (2\pi)^4 \delta^{(4)} (K - p_1 -p_2)\sum_{\text{pol.}} \vert{\cal M}\vert^2_{\gamma^{*}\to W^m W^p} \; ,
\end{aligned}
\eea
where $f_B(\omega(k)) = 1/(e^{\omega(k)/T} - 1)$ is the Bose-Einstein distribution function for the plasmon. 

For the longitudinal configuration, the thermally averaged decay rate is given by 
\bea
\label{eq:plasmonL}
n_{\gamma_\ell^*} \langle\Gamma\rangle_{\gamma_\ell^{*}\to W^p W^m} = \frac{g_D^2 \kappa_1^2}{96\pi^3}
 \int_0^{k_{\rm max}}  
k^2  dk Z_\ell(k) \omega_\ell(k) f_B (\omega_\ell(k)) \frac{\left(1 - 4 x_\ell \right)^{3/2} (1 + 3 x_\ell)}{x_\ell}  
\; ,
\eea
where $x_\ell =  m_{W^\prime}^2/ m^2_\ell$ 
and $k_{\rm max}$ is the maximum plasmon momentum given by~\cite{Braaten:1993jw}
\be
\label{eq:kmax}
k_{\rm max} = \left[ 
\frac{3}{v^2_*} \left( \frac{1}{2 v_*} \ln{\frac{1+ v_*}{1-v_*} - 1 }
\right)
\right]^{1/2} \omega_p \; .
\ee
In Eq.~(\ref{eq:kmax}), \( \omega_p \) is the plasma frequency, given by
\begin{equation}
\label{eq:plasmafreq}
\omega_p^2 = \frac{8 \alpha_{\rm EM}}{\pi} \int_0^\infty
dp \, \frac{p^2}{E} \left( 1 - \frac{1}{3} v^2 \right)
f_F ( E ) \; ,
\end{equation}
where \( v = p/E \), with \( E = \sqrt{p^2 + m_e^2} \), is the velocity of electrons or positrons, and \( f_F(E) = 1/(e^{E/T} + 1) \) is the Fermi-Dirac distribution. The parameter \( v_* \) is defined as
\begin{equation}
v_* = \frac{\omega_1}{\omega_p} \; ,
\end{equation}
where \( \omega_1 \) is given by
\begin{equation}
\omega_1^2 = \frac{8 \alpha_{\rm EM}}{\pi} \int_0^\infty
dp \, \frac{p^2}{E} v^2 \left( \frac{5}{3}  - v^2 \right) f_F ( E ) \; .
\end{equation}

For the transverse configuration, the thermally averaged decay rate is given by 
\bea
\label{eq:plasmonT}
\begin{aligned}
n_{\gamma_t^*} \langle\Gamma\rangle_{\gamma_t^{*}\to W^p W^m} &= \frac{g_D^2 \kappa_1^2 }{48\pi^3}
 \int_0^\infty 
k^2 dk \frac{Z_t(k) m_t^2 f_B (\omega_t(k)) }{\omega_t(k)} \frac{ \left(1 - 4 x_t \right)^{3/2} (1 + 3 x_t)} {x_t} \; ,
\end{aligned}
\eea
where $x_t =  m_{W^\prime}^2  / m^2_t$.

It may be interesting to extend the above analysis to include the dark plasmon decay process $Z^{\prime *} \to W^p W^m$ in the hidden heat bath as an additional freeze-in production channel. However, we expect this contribution to be minuscule due to the smallness of $g_D$ and therefore ignore it here.

\allowdisplaybreaks
\bibliographystyle{apsrev4-1}
\bibliography{refs}

\end{document}